\documentstyle[12pt,epsf]{article}

\input{psfig}

\newcommand{\nc}{\newcommand}
\def\frac#1#2{{\textstyle {#1 \over #2}}}

\nc{\beq}{\begin{equation}}
\nc{\eeq}{\end{equation}}
\nc{\beqa}{\begin{eqnarray}}
\nc{\eeqa}{\end{eqnarray}}
\nc{\lsim}{\begin{array}{c}\,\sim\vspace{-21pt}\\< \end{array}}
\nc{\gsim}{\begin{array}{c}\sim\vspace{-21pt}\\> \end{array}}
\nc{\eps}{\epsilon}
\nc{\s}{\sigma}
\nc{\veps}{\varepsilon}
\nc{\no}{\noindent}
\nc{\D}{\Delta}
\nc{\nn}{\nonumber}
\nc{\al}{\alpha}
\nc{\be}{\beta}
\nc{\ga}{\gamma}
\nc{\de}{\delta}

\begin{document}

\begin{center} 

\vskip .5 in
{\large \bf 
Coupled Minimal Models with and without Disorder}

\vskip .6 in

  {\bf  P. Simon}\footnote{simon@lpthe.jussieu.fr }
   \vskip 0.3 cm
 {\it   Laboratoire de Physique Th\'eorique et Hautes Energies  }\footnote{ 
 Unit\'e associ\'ee au CNRS URA 280}\\
 {\it  Universit\'es Pierre et Marie Curie Paris VI et Denis Diderot Paris 
VII}\\
{\it  2 pl. Jussieu, 75251 Paris cedex 05 }

  \vskip  1cm   
\end{center}
\vskip .5 in

 \begin{abstract}
We analyse in this article the critical behavior of $M$ $q_1$-state Potts models coupled to $N$ $q_2$-state Potts models ($q_1,q_2\in [2..4]$) with and without disorder. The technics we use are based on perturbed conformal theories. Calculations have been performed at two loops. We already find some interesting situations in the pure case for some peculiar values of $M$ and $N$ with new tricritical points. When adding weak disorder, the results we obtain tend to show that disorder makes the models decouple. Therefore,  no relations emerges, at a perturbation level, between for example the disordered $q_1\times q_2$-state  Potts 
model and the two  disordered $q_1$, $q_2$-state Potts models ($q_1\ne q_2$), despite their central charges are similar according to recent numerical 
investigations.

 \end{abstract}

\no 
PACS NUMBERS: 05.70.Jk; 64.60.Fr; 75.10.Hk; 75.40.Cx\\\no

\eject
\renewcommand{\thepage}{\arabic{page}}
\setcounter{equation}{0}
\noindent

\section{Introduction}
In order to understand the role of impurities and inhomogeneities in real physics systems, many statistical models with quenched randomness have been proposed. Therefore, the effect of randomness of continuous phase transitions have been of great interest for many years.
The main question is to determine if the randomness leaves unchanged the critical properties of the pure system. One prediction concerning models with random bonds comes from the Harris conjecture \cite{harris}, which states that bond randomness changes the values of the  critical exponents only if the specific-heat exponent $\alpha$ is positive.
The effects of bond randomness has been studied intensively in the  $2D$ Ising model (The Harris criterion does not enable to be conclusive about the relevance of disorder) first by 
Dotsenko and Dotsenko \cite{dots1}. They represent the near critical point by a Gross-Neveu model and found an  asymptotic behavior for the specific heat in $ln(\ln({1\over |t|})$ with $|t| $ the reduced temperature. This has been confirmed numerically \cite{num1}. For $2D$ random Potts model (the Harris criterion predicts here that  randomness is relevant and changes the critical behavior), conformal field theory techniques have been a powerful tool to compute in pertubation the shifts in the critical exponents \cite{ludwig,dots2}.
The numerical results have proved this approach to be accurate (see \cite{picco1} and
\cite{cardy1} for more recent results). All these results concern simple models having a second order phase transition in the pure case. \\
\no However, it has been realized recently, that a weak bond randomness can also have strong effects on $2D$ systems possessing a first order transition. Hence, following earlier work of Imry and Wortis \cite{imry}, Hui and Berker have shown with a phenomenological renormalisation group that bond randomness will induce a second order transition (a vanishing of the latent heat) in a system that would have undergone  a
first order one \cite{hui}. This result has been stated on a more rigorous level by Aizenman and Wehr \cite{aizen}. At this level, a question  naturally  arises concerning the universality class of this second order transition. In order to test these predictions, Cardy has studied a system presenting a
fluctuation driven first order transition, namely a transition which is expected to be continuous at the mean field level, but become first order when fluctuations are included.
Therefore, Cardy was able to show that the addition of weak randomnes on a system of $N$ coupled Ising models makes the system flow to $N$ decoupled Ising models \cite{cardy2}. This result has been extended to the case of $N$ ($N>2$) coupled Potts models by pujol \cite{pujol} (the case $N=2$ is quite peculiar and deserves a special attention \cite{pujol,simon1}).
The result was a non-Ising like second order transition. The universality class depends on the sign of the coupling between the models and therefore these systems seem to violate the Imry-Wortis argument \cite{imry}.

Naturally, the $q-$state Potts model with $q>4$ appears more suitable to test and study the effects of disorder, because the  first order transition is by now of mean field type.
The hint is that there is now analytical approach able to control the effect of disorder except at large $q$ where a mapping to the random field Ising model shows the absence of any latent heat in $2D$ \cite{cardy1}. For $q$ not too large, only numerical simulations
are able to check the Imry-Wortis arguments. Hence, Chen {\it et al.} \cite{chen} have investigated
the $8-$state Potts model using Monte Carlo simulations and confirmed the transition to be continuous but also found numerical values of the critical exponents consistent with those of the Ising model. Nevertheless, recent numerical studies of Cardy and Jacobsen \cite{cardy1} and of Picco \cite{picco2} are in a clear disagreement with the latter conclusion. The magnetization exponent ${\beta\over \nu}$ is found to vary continuously with $q$. Moreover, it has also been shown, using the powerful tool of conformal invariance combined with finite size scaling, that the values of the central charges of disordered $q-$state Potts models are related to one another by a factorization law
$c(q=q_1\times q_2)=c(q_1)+c(q_2)$ \cite{picco2,cardy1}. Therefore, the numerical measure of central charges can not distinguish between a disordered $q=q_1\times q_2$-state Potts model and two decoupled $q_1$ and $q_2$ disordered Potts models. As the only analytical results concern the behavior of similar coupled $q-$state Potts model ($q=2,3$) under weak disorder, it should be interesting to extend these  results to different disordered Potts models in order to compare the critical behavior of  coupled $q_1$ and $q_2$-state Potts models under weak randomness with the disordered $q_1\times q_2$-state Potts model.\\
\no
In order to bring some answers to this issue, the main goal of this paper is  therefore the study of the critical behavior of $M$ 
Ising models coupled to $N$ $q-$state Potts models ($q=3,4$) with disorder. Nevertheless, a special attention will be first paid to the pure case. \\
\no Indeed, it seems important to understand in a general context the critical behavior of different  minimal models notably coupled by their energy density. This question was recently adressed by Leclair {\it et al.} \cite{leclair} (see also \cite{vays}) in the context of integrable perturbations theories. Namely could the mixing of two  or more critical models create a new critical behavior ? Moreover, it might be a way to approach the critical behavior of some $3D$ systems. For example, when we couple one Ising model to $N$ $q$-state Potts model, we obtain using one loop calculations a non-trivial flow with new tricritical points \cite{simon2}. Nevertheless, the most puzzling situation concerns the unusual critical behavior of one Ising model coupled to an XY model, currently referred as Ising-XY in the literature \cite{isxy}. This model is expected to include in its range of parameters  a whole class
of systems with $Z_2\times U(1)$ symmetry like the $2D$ fully frustrated XY model \cite{ffxy} the $2D$ arrays of josephson junctions \cite{joseph}, the ANNNXY model \cite{simon3} and so on. The Ising-XY model have in its phase diagram a continuous critical line with simultaneous XY and Ising ordering where the critical exponent associated to the Ising magnetization varies continuously and differs significantly from the pure Ising one.
Therefore, along this critical line, we have two  models at their critical temperature coupled by their energy density and the numerical results indicate an unusual behavior. Those examples shows that an energy-energy coupling term may have drastic effects in certain situations. Therefore, before introducing weak disorder, we carefully study the critical behavior of $M$ 
Ising models  to $N$ $q-$state Potts models. We thus confirm, using two loops calculations, the existence of the two tricritical points quoted above. These results also extend when we consider $M$ $3-$state Potts models coupled to $N$ $4-$state Potts models.\\
\no
The paper is written in a self-contained way and   completes the work presented in \cite{simon1,simon2}. It is organized as follows. In section $2$, we analyse the critical behavior of $M$ Ising models coupled to $N$ Potts models. A specific attention is paid to the cases $M=1,2$. Near their critical points, these models are presented by perturbed conformal theory. Then, in section $3$, we study the effects of weak randomness on these models. We use the replica trick method. We show that disorder makes the models decouple. Therefore, our analysis confirms no apparent relations (at least perturbatively!) between the disordered $q_1\times q_2$-state Potts model and the corresponding coupled $q_1$, $q_2$ disordered Potts model,
despite their central charge are similar. Section $4$ contains a  summary of the results and some future
perspectives. All the integrals involved in the computation of the beta functions, the  renormalisations of the spin and energy
operators  are worked out explicitly in the appendices.

\section{Coupled minimal models without disorder}
\setcounter{equation}{0}
\subsection{The model}
The model we indend to study in this section consists of a  superposition of $M$ Ising  and $N$ $q$-state Potts models coupled by their energy density. The system possess a $Z_2^M\times Z_q^N$ symmetry.
The Hamiltonian has the following form
\beq
\label{ham}
H=\sum\limits_{a=1}^M H_{1}^a+\sum\limits_{b=M+1}^{M+N} H_{2}^b- 
\sum\limits_{c,d} g_{cd} \int d^2 z~\varepsilon^c(z)\varepsilon^d(z)~,
\eeq
where $c,d$ belong to $[1,\cdots,M+N]$. $H_1^a$ and $H_{2}^b$ represent respectively one
pure Ising model and one pure $q-$state Potts model at their respective critical temperature. $\varepsilon^c$ corresponds to the energy operator of the $c^{th}$ model (either Ising or Potts depending on the value of $c$). Therefore, the partition function reads as
\beq
\label{z}
Z = \prod_{a=1}^m Tr_{a,s_{i,a}} \prod_{b=M+1}^{M+N} 
Tr_{b,s_{i,b}}e^{-\displaystyle\sum_{a=1}^m
S_{0,a} -\sum_{b=M+1}^{M+N}
S_{0,b}+  \int \sum\limits_{c\ne d}g_{cd}\displaystyle
\varepsilon^c(z)\varepsilon^d(z) d^2z ~.
}
\eeq
This model can be described as a  pure conformal field theory perturbed by the interaction term  which is quadratic in the $\veps$ operators. The infrared behavior of the coupling constants of the complete model ($g_{ij}$ and also the different masses that have not been included in (\ref{ham})) can be analysed perturbatively. In a similar way, a correlation function like
 $<O(0)O(R)>$, where $O$ is some local operator, can be expanded 
perturbatively like:
$$
<O(0)O(R)> = <O(0)O(R)>_0+<S_IO(0)O(R)>_0+{1\over2}<S_I^2O(0)O(R)>_0+\cdots
$$
where $<>_0$ means the expectation value taken with respect to the bare action and 
\beq
S_I =
\int H_I(z) d^2z =  \int \displaystyle\sum_{c\not=d}g_{cd}
\varepsilon^c(z)\varepsilon^d(z) d^2z~,
\eeq
the interaction term.

\subsection{Coulomb gas formulation}
In order to compute the integrals involved in the calculus of correlations functions, we used the Coulomb-gas representation of a conformal field theory \cite{fateev}. In this representation, the central charge is written as $c={1\over 
2}+\epsilon$, where $\epsilon$ will be used as a short distance regulator for 
the integrals involved in correlation functions calculations. In our calculations, we are obliged to introduce two UV regulators $\eps_1,\eps_2$, since we mix two different minimal models. In addition, we also use an I.R. cut -off $r$ which is useful to derive the RG equations. Then the limit 
$\eps_1\to 0$ and  a finite value for $\eps_2$  corresponds to  Ising models coupled to $q-$state Potts models. The integrals will thus be expressed as finite series in $\eps_1,\eps_2$ with coefficients depending on $r$.\\
In the Coulomb gas representation, the central charge $c$ will be
characterized in the following by the parameter $\alpha^2_+ = {2p\over2p-1}
={4\over 3} + \epsilon $ with
\beqa
c=1-24\alpha_0^2~~&;&~~\alpha_{\pm}=\alpha_0\pm\sqrt{\alpha_0^2+1};\\
\alpha_+\alpha_-&=&-1 .\nonumber
\eeqa
For the pure Ising model   $\alpha_+^2={4\over 3}$ and $c={1\over 2}$ while for 
the $3-$states Potts model  $\alpha_+^2={6\over 5}$, $c={4\over 5}$ and 
$\eps=-{2\over 15}$. The $4-$states Potts model can be considered as the limiting case of this  perturbative scheme \cite{dots2}, it has c=1 and $\eps=-{1\over 3}$. The vertex operators are defined by 
$V_{nm}(x)=e^{i\alpha_{nm}\Phi(x)}$ where $\Phi(x)$ is a free scalar field and 
the $\alpha_{nm}$ are defined by
\beq
\alpha_{nm} = {1\over 2}(1-n)\alpha_-+{1\over 2}(1-m)\alpha_+~.
\eeq
The conformal dimension of the operator $V_{nm}(x)$ is 
$\Delta_{nm}=-\alpha_{\overline{nm}}\alpha_{nm}$ with
\beq
\alpha_{\overline{nm}}=2\alpha_0-\alpha_{nm}={1\over 2}(1+n)\alpha_-+{1\over 
2}(1+m)\alpha_+~.
\eeq
The spin field $\sigma$ can be represented by the vertex operator $V_{p,p-1}$ 
whereas $V_{1,2}$ corresponds to the energy operator $\veps$. Note that in the 
Ising case the $\sigma$ operator could also be represented by the $V_{2,1}$. Therefore, the spin operator can be represented in the general case (Ising or Potts) by $V_{k,k-1}$ where $k={2+3\lambda\eps\over 1+3\eps}$. We thus have $\lambda=2$ for $V_{21}$ and $\lambda={1\over 2}$
 for $V_{p,p-1}$. All these notations generalise in a straightforward way when we mix two minimal models since we introduce two UV regulators $\eps_1$ and $\eps_2$.

\subsection{Renormalisation group equations}
In this section, we deal with the renormalisation of the $\s_i$ and $\veps_i$ ($i=1,2$) operators. Nevertheless, we first need to compute
 the renormalisation of the couplings constants $g_{ij}$.

\subsubsection{Beta functions}
When we have only one coupling constant $g_0$, 
its renormalisation is determined
directly by a perturbative computation. $g$ is also
given by the O.A. producing
\beq
g_0 \int \displaystyle\varepsilon_a(z)\varepsilon_b(z) d^2z
+ {1\over 2} \left(g_0 \int \displaystyle
\varepsilon_a(z)\varepsilon_b(z) d^2z\right)^2 +\cdots= g\int 
\displaystyle\varepsilon_a(z)\varepsilon_b(z) d^2z~,
\eeq
with $g=g_0 + A_2 g_0^2  + \cdots $ where $A_2$ comes from the contraction
\beq
{1\over 2} \int \displaystyle 
\varepsilon_a(z)\varepsilon_b(z) d^2z \int \displaystyle 
\varepsilon_c(z)\varepsilon_d(z) d^2z = A_2 \int
\displaystyle\  \varepsilon_a(z)\varepsilon_b(z) d^2z 
\eeq
\noindent
This  procedure generalizes to two loops and more. In this article, we restrict to two loops calculations.\\
All this formulation can be extended in a straighforward way to our case. We suppose
that $g_{cd}=g_1$ for $c,d \in [1,M]$, $g_{cd}=g_2$ for $c,d \in [M+1,M+N]$ and
finally  $g_{cd}=g_{12}$ for $c\in [1..M], d\in[M+1,M+N]$ or vice versa. The technical details are explained in the appendix A.
We find
\beqa
\label{renor1}
g_1(r)&=&r^{-3\eps_1}\left(g_1^0-4\pi(M-2)(g_1^0)^2 {r^{-3\eps_1}\over 3\eps_1} -4\pi N (g_{12}^0)^2  {r^{-3\eps_2}\over 3\eps_2} \right.\nn\\&&~~~~~~~~
+8\pi^2(M-2)(g_1^0)^3 {r^{-6\eps_1}\over 3\eps_1}\left[1+ {2(M-2)\over 3\eps_1}\right]\nn\\&&~~~~~~~~+32\pi^2 N(M-2) g_1^0(g_{12}^0)^2 {r^{-3(\eps_1+\eps_2)}\over 3(\eps_1+\eps_2)}\left[{1\over 3\eps_1}+{1\over 3\eps_2}\right]\\&&~~~~~~~~ +\left.16\pi^2 N g_1^0(g_{12}^0)^2 {r^{-3(\eps_1+\eps_2)}\over 3(\eps_1+\eps_2)}\left[1+{2\over 3\eps_1}\right] +16\pi^2 N(N-1)g_2^0(g_{12}^0)^2 {r^{-6\eps_2}\over 9\eps_2^2}~~\right)\nn
\eeqa
\beqa
\label{renor2}
g_2(r)&=&r^{-3\eps_2}\left(g_2^0-4\pi(N-2)(g_2^0)^2 {r^{-3\eps_2}\over 3\eps_2} -4\pi M (g_{12}^0)^2  {r^{-3\eps_1}\over 3\eps_1}\right.\nn\\&&~~~~~~~~
+8\pi^2(N-2)(g_2^0)^3 {r^{-6\eps_2}\over 3\eps_2}\left[1+ {2(N-2)\over 3\eps_2}\right]\nn\\&&~~~~~~~~+32\left.\pi^2 M(N-2) g_2^0(g_{12}^0)^2 {r^{-3(\eps_1+\eps_2)}\over 3(\eps_1+\eps_2)}\left[{1\over 3\eps_1}+{1\over 3\eps_2}\right]\right.\\&&~~~~~~~~ +\left.16\pi^2 M g_2^0(g_{12}^0)^2 {r^{-3(\eps_1+\eps_2)}\over 3(\eps_1+\eps_2)}\left[1+{2\over 3\eps_2}\right] +16\pi^2 M(M-1)g_1^0(g_{12}^0)^2 {r^{-6\eps_1}\over 9\eps_1^2}~~\right)\nn
\eeqa
\beqa
\label{renor3}
g_{12}(r)&=& r^{-3(\eps_1+\eps_2)}\left( g_{12}^0-4\pi(M-1)g_1^0g_{12}^0  {r^{-3\eps_1}\over 3\eps_1} -4\pi(N-1)g_2^0 g_{12}^0  {r^{-3\eps_2}\over 3\eps_2}\right.\nn\\ 
&&~~+4\pi^2 (M-1)(g_1^0)^2g_{12}^0  {r^{-6\eps_1}\over 3\eps_1}\left[1+{2(M-1)\over 3\eps_1}\right]\nn\\&&~~+4\pi^2 (N-1)(g_2^0)^2g_{12}^0  {r^{-6\eps_2}\over 3\eps_2}\left[1+{2(N-1)\over 3\eps_2}\right]\nn\\&&~~+\left.16\pi^2 (N-1)(M-1)g_1^0g_2^0g_{12}^0 {r^{-3(\eps_1+\eps_2)}\over 3(\eps_1+\eps_2)} \left[{1\over 3\eps_1}+{1\over 3\eps_2}\right]\right.\\&&~~
+\left.16\pi^2 (N-1)(M-1) (g_{12}^0)^3{r^{-3(\eps_1+\eps_2)}\over 3(\eps_1+\eps_2)}  \left[{1\over 3\eps_1}+{1\over 3\eps_2}\right] \right.\nn\\&&~~
+\left.8\pi^2 (g_{12}^0)^3{r^{-3(\eps_1+\eps_2)}\over 3(\eps_1+\eps_2)}\left[M+N-2+{2(M-1)\over 3\eps_1}+{2(N-1)\over 3\eps_2}\right]~~\right)\nn
\eeqa
We have multiplied each equations by a power of $r$ in order to have dimensionless coupling constants $g_i(r)$. From these expressions, it is then easy to obtain the three beta fonctions associated to $g_1,g_2,g_{12}$. We obtain the following result
\beqa
\label{beta1}
\beta_1={d g_1\over d(ln l)}&=& \eps_1~g_1+ (M-2)~g_1^2+ N~ g_{12}^2-(M-2)~g_1^3-N~ g_1g_{12}^2+o(g^3),
\\
\label{beta2}
\beta_2={d g_2\over d(ln l)}&=& \eps_2~ g_2+(N-2)~g_2^2+4\pi M~
g_{12}^2-(N-2)~g_2^3-M~g_2g_{12}^2+o(g^3),\\
\label{beta3}
\beta_{12}={d g_{12}\over d(ln l)}&=& {\eps_1+\eps_2\over 2}~ 
g_{12}+(M-1)~g_1g_{12}+ (N-1)~g_2g_{12}-{1\over 2}(M+N-2)~g_{12}^3\nn\\&&-{1\over 2}(M-1)~g_1^2g_{12}-{1\over 2}(N-1)~g_2^2g_{12}+o(g^3)~.
\eeqa
In the above equations, we have made for simplicity the changements $ g\to 4\pi g$ and $\eps_i\to -3\eps_i$. If we consider the case of $M$ Ising models coupled to $N$ $3-$state Potts models, we take the limits $\eps_1\to 0$ and $\eps_2\to {2\over 5}$. The fixed point structure is studied in a further section.

\subsubsection{Renormalisation of $\veps_i$} 
Under the action of the perturbation term, the operators  $\veps_i$ will be renormalised into $\veps'_i$. It means that we have to compute the $2\times 2$ matrix $Z_{\veps}$ defined by
$\tilde{\veps}'= Z_{\veps}\tilde{\veps}$,
where $\tilde{\veps}'$  are two components vectors.
Indeed, since the interaction term mixes different energy operators, the matrix $Z_{\veps}$
contains off-diagonal terms. In order to compute these matrix function, we add source terms to $\tilde{\veps}$  in the action, namely mass terms. Therefore, the renormalisation of the operator $\tilde{\veps}$ is equivalent to the renormalisation of the mass term $\tilde{m}$ defined by
\beq
\tilde{m}~\tilde{E}=\tilde{m^0}~Z_{\veps}~\tilde{E},
\eeq with $$\tilde{E}=\left(\int\sum\limits_{a=1}^M\veps_1^a(z) d^2z,\int\sum\limits_{b=M+1}^{M+N}\veps_2^b(z) d^2z\right).$$
As for the computation of the beta fonctions, we can compute in perturbation the renormalisation of $\tilde{m}$. The details of the computation are presented in appendix B. We thus have
\beqa
\label{mas1}
m_1(r)r^{-1+{3\over 2} \eps_1}&=& m_1^0\left( 1- 4\pi(M-1)g_1^0 {r^{-3\eps_1}\over 3\eps_1}
+4\pi^2 (M-1)(g_1^0)^2  {r^{-6\eps_1}\over 3\eps_1}\left[1+{4M-6\over 3\eps_1}\right]\right.\nn\\
&&+\left.8\pi^2 N(g_{12}^0)^2{r^{-3(\eps_1+\eps_2)}\over 3(\eps_1+\eps_2)}\left[1+{2M\over 3\eps_1}+{2(M-1)\over 3\eps_2}\right]~~\right)\\&&
+ m_2^0\left(-4\pi N g_{12}^0 {r^{-3\eps_2}\over 3\eps_2}+8\pi^2 N(M-1)g_1^0g_{12}^0 {r^{-3(\eps_1+\eps_2)}\over 9\eps_1\eps_2}\right.\nn\\&& +\left.8\pi^2 N(N-1)g_2^0g_{12}^0 {r^{-6\eps_2}\over 9\eps_2^2}\right)\nn
\eeqa
\beqa
\label{mas2}
m_2(r)r^{-1+{3\over 2} \eps_2}&=& m_2^0\left( 1- 4\pi(N-1)g_2^0 {r^{-3\eps_2}\over 3\eps_2}
+4\pi^2 (N-1)(g_2^0)^2  {r^{-6\eps_2}\over 3\eps_2}\left[1+{4N-6\over 3\eps_2}\right]\right.\nn\\
&&+\left.8\pi^2 M(g_{12}^0)^2{r^{-3(\eps_1+\eps_2)}\over 3(\eps_1+\eps_2)}\left[1+{2N\over 3\eps_2}+{2(N-1)\over 3\eps_1}\right]~~\right)\\&&
+ m_1^0\left(-4\pi M g_{12}^0 {r^{-3\eps_1}\over 3\eps_1}+8\pi^2 M(N-1)g_2^0g_{12}^0 {r^{-3(\eps_1+\eps_2)}\over 9\eps_1\eps_2}\right.\nn\\&& +\left.8\pi^2 M(M-1)g_1^0g_{12}^0 {r^{-6\eps_1}\over 9\eps_1^2}\right)\nn
\eeqa
Here again, we have multiplied the two renormalised masses by a power of $r$ in order to obtain a dimensionless coupling constant. From these equations and the renormalisations of the coupling constants (\ref{renor1},\ref{renor2},\ref{renor3}), we obtain the RG equations at two loops associated with the matrix $Z_{\veps}$ (we have again used the normalisations $ g\to 4\pi g$)
\beq
\label{masse}
{d\log(Z_{\veps})\over \log(r)}=\left(\begin{array}{cc}
(M-1)g_1-{1\over 2}(M-1)g_1^2-{1\over 2} N g_{12}^2& N g_{12}\\
 M g_{12}&(N-1)g_1-{1\over 2}(M-1)g_1^2-{1\over 2} M g_{12}^2\end{array}\right)~~~
\eeq
The logarithm of the matrix has been defined by its series expansion. We clearly see that both energy operators of the two models are mixed under the perturbation term. Therefore, the bare operators $\veps_i$ are not the true energy eigen-operators of the problem.

\subsubsection{Renormalisation of $\s_i$}
In a similar way, the spin operators $\s_i$ will be renormalised in $\s_i'$. We can thus define the fonctions $Z_{\s_1},~Z_{\s_2}$ by $\s_i'=Z_{\s_i}\s_i$ with $i=1,2$ (the interaction term does not mix the spin operators). A convenient way to compute these functions is to introduce two magnetic field $h_i^0$ which couple to the spins $\s_i$. Then, we have to calculate the effect of the perturbation term 
$\sum\limits_{c,d} g_{cd} \int d^2 z~\varepsilon^c(z)\varepsilon^d(z)$ on the coupling terms $\sum\limits_{i=1,2}h_i^0\sum\limits_{a_i=1}^{M+N} \s_i^{a_i}(z) d^2z$. More precisely, the corrections to the spin operators will come from the operator algebra between the $\veps$ and $\s$ operators. Therefore, as in the previous subsections, we can compute these corrections perturbatively in power of $g_{ij}$ up to third order. The computation is detailed in the appendix C. We find
\beqa
\label{champs1}
h_1(r)r^{-{15\over 8}-a(\eps_1)}&=&h_1^0\left(1+ (M-1)(g_1^0)^2\pi^2{r^{-6\eps_1}\over 2}\left[1+{4\over 3}(2-\lambda_1){\cal R}\right]\right.\nn\\&&
+{3N\over 2} (g_{12}^0)^2\pi^2 {r^{-3(\eps_1+\eps_2)}\over 3(\eps_1+\eps_2) }\left[\eps_2+\eps_1(1+{8\over 3}(2-\lambda_1){\cal R})\right]\nn\\&&-12(M-1)(M-2)(g_1^0)^3\pi^3 {r^{-9\eps_1}\over 9\eps_1}\left[1+{8\over 9}(2-\lambda_1){\cal R}\right]\\&&  -8N(N-1)\pi^3 (g_{12}^0)^2 (g_2^0) {r^{-3\eps_1-6\eps_2}\over 3\eps_1+6\eps_2 }\left[1+{\eps_1\over 2\eps_2}(1+{8\over 3}(2-\lambda_1){\cal R})\right]\nn\\&&
-\left.8N(M-1)\pi^3 (g_{12}^0)^2g_1^0 {r^{-6\eps_1-3\eps_2}\over 6\eps_1+3\eps_2 }\left[(1+{\eps_1\over \eps_2})(1+{4\over 3}(2-\lambda_1){\cal R})+{1\over 2}(1+{\eps_2\over \eps_1}) \right] ~~\right)\nn,
\eeqa
and in a similar way by exchanging $(1,M) \leftrightarrow (2,N)$
\beqa
\label{champs2}
h_2(r)r^{-{15\over 8}-a(\eps_2)}&=&h_2^0\left(1+ (N-1)(g_2^0)^2\pi^2{r^{-6\eps_2}\over 2}\left[1+{4\over 3}(2-\lambda_2){\cal R}\right]\right.\nn\\&&
+{3M\over 2} (g_{12}^0)^2\pi^2 {r^{-3(\eps_1+\eps_2)}\over 3(\eps_1+\eps_2)}
\left[\eps_1+\eps_2(1+{8\over 3}(2-\lambda_2){\cal R})\right]
\nn\\&&-12(N-1)(N-2)(g_2^0)^3\pi^3 {r^{-9\eps_2}\over 9\eps_2}\left[1+{8\over 9}(2-\lambda_2){\cal R}\right]\\&&  -8M(M-1)\pi^3 (g_{12}^0)^2 (g_1^0) {r^{-6\eps_1-3\eps_2}\over 6\eps_1+3\eps_2 }\left[1+{\eps_2\over 2\eps_1}(1+{8\over 3}(2-\lambda_2){\cal R})\right]\nn\\&&
-\left.8M(N-1)\pi^3 (g_{12}^0)^2g_2^0 {r^{-3\eps_1-6\eps_2}\over 3\eps_1+6\eps_2 }\left[(1+{\eps_2\over \eps_1})(1+{4\over 3}(2-\lambda_2){\cal R})+{1\over 2}(1+{\eps_1\over \eps_2})\right] ~~\right)\nn,
\eeqa
where we have defined
 $${\cal R}={\Gamma^2(-{2\over 3})\Gamma^2({1\over 6})
\over \Gamma^2(-{1\over 3})\Gamma^2(-{1\over 6})}.$$
As usual, the multiplicative term $r^{-{15\over 8}-a(\eps_i)}$ in front of $h_i(r)$
is introduced in order to deal with dimensionless parameters. The function $a(\eps_i)$ is a function of $\eps_i$ depending on which representation of the spin field we take in the Coulomb gas picture (see section 2.2). Now, we deduce from the above equations and from (\ref{renor1},\ref{renor2},\ref{renor3}), the RG equations associated to $Z_{\s_1},~Z_{\s_2}$ (with $\eps_i\to -3\eps_i$)
\beqa
\label{sigma1}
{d\log(Z_{\s_1})\over d\log(r)} &=&(M-1)g_1^2(r)\pi^2\eps_1\left[1+{4\over 3}(2-\lambda_1){\cal R}\right] \nn\\&&+Ng_{12}^2(r){\pi^2\over 2}\left[\eps_2+\eps_1(1+{8\over 3}(2-\lambda_1){\cal R})\right]\nn\\&&+4(M-1)(M-2)\pi^3g_1^3(r)+4N(N-1)\pi^3g_{12}^2(r)g_2(r)\\&&+4N(M-1)\pi^3g_{12}^2(r)g_1(r)\nn
\eeqa
\beqa
\label{sigma2}
{d\log(Z_{\s_2})\over d\log(r)} &=&(N-1)g_2^2(r)\pi^2\eps_2 \left[1+{4\over 3}(2-\lambda_2){\cal R}\right]\nn\\&& +Mg_{12}^2(r){\pi^2\over 2}\left[\eps_1+\eps_2(1+{8\over 3}(2-\lambda_2){\cal R})\right]\nn\\&&+4(N-1)(N-2)\pi^3g_1^3(r)+4M(M-1)\pi^3g_{12}^2(r)g_1(r)\\&&+4M(N-1)\pi^3g_{12}^2(r)g_2(r)\nn
\eeqa
\subsection{Fixed point structure}
In this section, we analyse the fixed points structure associated with the beta fonctions defined in (\ref{beta1}-\ref{beta3}). We focus in this paper on the case of Ising models coupled to $3$ or $4$-states Potts models. Indeed, the $3-$state Potts model corresponds to $\eps={2\over 5}$ in our reduced normalisation whereas the $4-$state Potts model corresponds to $\eps=1$. One can wonder about the validity of the perturbation expansion for the latter case. In fact, the beta fonctions have still fixed points order by order. And surprisingly, the approximation still remains quite good qualitatively. For example, if we compare the magnetic exponent of the disordered $4$-state Potts model predicted by the theory with numerical results, we find an error less than $5\%$ \cite{cardy1}.

 We now take the limit $\eps_1\to 0$ and $\eps_2\to \eps$ in the equations (\ref{beta1}-\ref{beta3}). They reduce to
\beqa
\label{beta}
\beta_1={d g_1\over d(ln l)}&=& (M-2)~g_1^2+ N~ g_{12}^2-(M-2)~g_1^3-N~ g_1g_{12}^2,
\nn\\
\beta_2={d g_2\over d(ln l)}&=& \eps~ g_2+(N-2)~g_2^2+4\pi M~
g_{12}^2-(N-2)~g_2^3-M~g_2g_{12}^2,\\
\label{g12}
\beta_{12}={d g_{12}\over d(ln l)}&=& {\eps\over 2}~ 
g_{12}+(M-1)~g_1g_{12}+ (N-1)~g_2g_{12}-{1\over 2}(M+N-2)~g_{12}^3\nn\\&&-{1\over 2}(M-1)~g_1^2g_{12}-{1\over 2}(N-1)~g_2^2g_{12}\nn~.
\eeqa
We consider in the following different peculiar values of $(M,N)$.

\subsubsection{Case $M=1$}
It corresponds to one Ising model coupled to one or several Potts model.\\
$\bullet$ Let us first analyse the case of one Ising model couple to one $q$-state Potts model. Then, the equation (\ref{beta3}) reduces to
\beq
\beta_{12}={d g_{12}\over d(ln l)}={\eps\over 2}~ g_{12}+o(g_{12}^3)
\eeq
We clearly see that the system is then driven in a strong coupling regime indicating probably a first order transition though a non-perturbative fixed point cannot be ruled out. A similar situation has been encountered when one couple two $q$-state Potts model \cite{pujol}. Moreover, in the latter case, it can be proved exactly that we have a mass gap generation \cite{vays}. Consequently,  when one superposes a critical Ising model to a critical $3$ or $4$ states potts model, no new critical behaviour is found in  pertubation when coupling both models by their energy density. Of course, new non-perturbative fixed points cannot be excluded by this method. Therefore, It suggests (perturbatively) that no line with simultaneous disordering of the 
Ising and Potts order parameter survives in the phase diagram of this model contrary to those of the FFXY model \cite{isxy}. Similar conclusions can be drawn with two different Potts models.  \\
$\bullet$
We now consider the case of one Ising model coupled to $N>1$ Potts models. We take $M=1$ in the equations (\ref{beta}). We find the following fixed points:
\beqa
\label{ptfix1}
&&g_1^*=g_2^*=g_{12}^*=0\\
\label{ptfix2}
&&g_1^*=g_{12}^*=0;~g_2^*={-\eps\over (N-2)}+{\eps^2\over (N-2)^2}~~ if~~ n>2\\
\label{ptfix3}
&&\left\{ \begin{array}{c}
g_1^*={\eps\tau_1 N\over 2(N-1)}-{\tau_1\eps^2(3N-1)\over 8(N-1)^2}\\~\\
g_2^*={-\eps\over 2(N-1)}+{(N+1)\eps^2\over 8(N-1)^2}\\~\\
g_{12}^*={\eps\tau_2\sqrt{N}\over 2(N-1)}-{\tau_2\eps^2(3N-1)\over 8\sqrt{N}(N-1)^2}
\end{array}\right.
\eeqa
with $\tau_1,\tau_2=\pm1$. There are four new fixed points (\ref{ptfix3}). We can study the stability of each of these fixed points by re-expressing  the solutions of (\ref{beta}) around these fixed points. Namely, we write $g_i=g_i^*+\delta g_i$ and keep the smallest order in $\eps$. This gives a linear system $\delta\dot{g}=A~\delta g$, with $\delta g$ a three component vector and $A$ a real $3\times3$ matrix. Therefore, the informations about the stability of each fixed points is obtained by calculating the eigenvalues of $A$ in each cases. By the way, we can see that the trivial fixed point (\ref{ptfix1}) is unstable, and that (\ref{ptfix2}) if stable in the $(g_2,g_{12})$ plane and marginal in the $g_1$ direction. It corresponds to the infrared fixed point of $N$ coupled Potts models.
 Concerning the other four fixed points (\ref{ptfix3}), two are unstable in two directions (when $\tau_1=-1$) and the order two ($\tau_1=+1$) are hyperbolic fixed points, namely unstable only in one direction. They constitute new tricritical points.
The local stable plane is defined by the vectors $$(1,0,0)~~;~~(0,{\tau_2\sqrt{2n(n-1)+1}-1\over \sqrt{n}(n-1)},1)$$ in the $(g_1,g_2,g_{12})$ space. A projection of the flow in the
$(g_{12},g_2)$ plane has been given in Figure 1 for the case $N=3$. The flow is symmetric along the $g_2$ axis. If initially $g_2^0>0$ then the flow is always driven in a strong coupling regime indicating mass generation as for the simple $N$-colored Potts model \cite{pujol}. Nevertheless, if $g_2^0<0$ and small, the trajectory will be first attracted by one of the tricritical points $T_1$ or $T_2$ (depending on the sign of $g_{12}^0$) and then flow toward the stable fixed point of the $N$-colored Potts model. The case $N=2$ is quite different since the fixed point (\ref{ptfix2}) is rejected to the infinity, therefore the system always flow in a strong coupling regime except on the stable surface in the $(g_1,g_2,g_{12})$ space.  It is a reminiscence of the exactly integrable $2$-colored Potts model which is always massive \cite{vays}.\\

If we consider the case of one $3$-state Potts model coupled to $N$ $4$-state models, we can derive exactly similar conclusions. The four tricritical points are now defined by (at first order in $\eps_1,\eps_2$)
\beq
\left\{ \begin{array}{c}
g_1^*={1\over 2}\left[\eps_1+\tau_1\sqrt{\eps_1^2+{N\over (N-1)^2}(\eps_1+\eps_2)(N(\eps_2-\eps_1)+2\eps_1)} \right]\\~\\
g_2^*={-(\eps_1+\eps_2)\over 2(N-1)}\\~\\
g_{12}^*={\tau_2\over 2(N-1)}\sqrt{(\eps_1+\eps_2))(N(\eps_2-\eps_1)+2\eps_1)}
\end{array}\right.
\eeq
We find a similar structure of fixed points as the previous case, namely two are stable in one direction, two are stable in two directions.

\vskip 0.2cm
In order to better characterise these fixed points, we can compute the corrections to the critical exponents of the pure case. The energy operators are completly mixed at the tricritical points and are therefore not the most appropriate variables, especially for a numerical simulation. Nevertheless, we can compute the renormalisation of the critical exponents of the spin-spin correlation functions. Hence, we use the Callan-Symanzik equations which give the form of the correlation functions.
 For one coupling constant $g$, we have
\beq
\label{exp}
<\sigma_i(0)\sigma_i(sL)>_{a,g_0} =
e^{2\int\limits_{g_0}^{g(s)}{\gamma_{\s_i}(g)\over \beta(g)}
dg}s^{-2\Delta_{\sigma_i}} <\sigma_i(0)\sigma_i(L)>_{r,g(s)}
\eeq
where we used the notation :
$
{dln(Z_\sigma) \over dln(r)} = \gamma_{\s}(g)
$
; $ g(a)= g_0 $ and $g(s)$ is defined by
$$\int\limits_{g_0}^{g(s)} \beta(g) dg = \log(s)~~;~ r=sa,$$ with $a$  the
lattice cut-off scale. $L$ is an arbitrary scale which can be fixed to one lattice spacing $a$ of a true statistical model. The dependence in $s$ of the term $ <\sigma_i(0)\sigma_i(L)>_{r,g(s)}$ is thus negligible and can be considered as a constant. We want to compute the corrections to the critical exponents close to the infrared tricritical points defined in (\ref{ptfix3}). Hence, the integral is dominated by the region $g\sim g^*$ and we have $\int\limits_{g_0}^{g(s)}{\gamma_{\s}(g)\over \beta(g)}
dg\sim\gamma_{s}(g^*)\log(s)$. We thus obtain
\beq
<\sigma(0) \sigma(s)>_{g_0} \sim
s^{-(2\Delta_{\sigma}-2\gamma_{\s}(g^*)) }
\eeq
Therefore, $-2\gamma_{\s_i}(g^*)$ corresponds to the correction of the magnetic critical exponent of the operator $\s_i$.  Using the equations (\ref{sigma1},~\ref{sigma2}) with $\eps_1=0,~\eps_2=\eps,~ M=1$, we find
\beqa
\gamma_{\s_1}(g^*)&=&N{\pi^2\over 2}\eps (g_{12}^*)^2+4N(N-1)\pi^3(g_{12}^*)^2g_2^* 
=0+O(\eps^4)\\
\gamma_{\s_2}(g^*)&=&(N-1)(g_2^*)^2\pi^2\eps\left[1+2{\cal R}\right]+(g_{12}^*)^2{\pi^2\over 2}\eps\left[1+4{\cal R}\right]\nonumber\\&&+4(N-1)(N-2)\pi^3(g_2^*)^3+4(N-1)\pi^3(g_{12}^*)^2g_2^*\nonumber\\
&=&{N\eps^3\over 128(N-1)^2}+{(2N-1)\eps^3\over 32(N-1)^2}{\cal R}+O(\eps^4),
\eeqa
with $\eps={2\over 5}$ for the $3$-state Potts model and $\eps=1$ for the $4$-state Potts model. We thus find the surprising result that there is no renormalisation of the critical exponent $\Delta_{\s_1}$ associated to the Ising spin at third order in $\eps$! Yet, the corrections to   $\Delta_{\s_2}$ are non zero and different from those of the infrared fixed point of $N$ coupled Potts models \cite{pujol}. Therefore, it sems that the physics at these tricritical points is non-trivial. It should be interesting to analyse and test numerically these predictions despite the corrections to the magnetization are very small.
Before considering the  other cases, we mention one remark.  Suppose we have a theory described by  one of the tricritical fixed points $T_i$, $i=1,2$  (see Figure 1), then a small perturbation can send the flow   to the fixed point $D$ characterized by decoupled Potts and Ising models. The central charge at this point  $c_{D}$ is thus the sum of the central charges of each model. Therefore according to Zamolodchikov c-theorem \cite{zamolod}, the central charge at $T_i$ ($c_{T_i}$) verifies $c_{T_i}\ge c_{D}$. A similar scenario could be imagined for the FFXY model with a new tricritical point inducing  a strong cross-over regime. Such a scenario  would justify strong finite size effects and $c_{FFXY}>{3\over 2}$ as it has been established numerically \cite{isxy,ffxy}.\\

\subsubsection{Case $M>1$}
We analyse in this subsection different peculiar cases and the general one.\\
$\bullet$ Let first begin with $M=2,N=1$.
The RG equations reduce in this case to

\beqa
\beta_1={d g_1\over d(ln l)}&=&   g_{12}^2- g_1g_{12}^2
\nn\\
\beta_2={d g_2\over d(ln l)}&=& \eps g_2-~g_2^2+2
g_{12}^2+g_2^3-2g_2g_{12}^2\\
\beta_{12}={d g_{12}\over d(ln l)}&=& {\eps\over 2}~ 
g_{12}+g_1g_{12}-{1\over 2}g_{12}^3-{1\over 2}g_1^2g_{12}~.\nn
\eeqa

The fixed points are
\beq
g_1^*=?~;~g_2^*=\left\{\begin{array}{c} \eps+\eps^2\\0\end{array}~;~g_{12}^*=0~~\right.
\eeq
We recover the line of fixed point parametrised by $g_1^*$ which corresponds to the Ashkin-Teller model model \cite{cardy2}. 
We can study (perturbatively) the stability of this line of fixed point especially according to the variable $g_{12}$ (around $g_2^*=\eps+\eps^2$ corresponding to the fixed point of the Potts model). We have
\beq
{d\delta g_{12}\over \log l}=({\eps\over 2}+g_1^*-{1\over 2} (g_1^*)^2)\delta g_{12}
\eeq
We thus see that the line of fixed point is stable only for $g_1^*< -{\eps\over 2}+{\eps^2\over 8}$. We have drawn on Figure 2 a shematic representation of the flow in the $(g_1,g_{12})$ plane). The point P$(g_1^*= -{\eps\over 2}+{\eps^2\over 8};g_2^*=\eps+\eps^2; g_{12}^*=0)$ separates the line parametrized by $g_1^*$ in two parts, a stable one and an unstable one. At the fixed point P, we can compute the corrections to the spin and energy critical exponents.
We find on one hand  no correction to the Ising magnetization and on the other hand $$2\Delta'_{\veps_1}(P)=2\Delta_{\veps}-
\gamma_{\veps_1}(P)=2\Delta_{\veps_1}+\eps+o(\eps^2).$$\\

This result may be checked numerically.

$\bullet$ If we now consider the case $M=2,N=2$, we can draw similar conclusions than the previous case except that the fixed point $g_2^*=\eps+\eps^2$ has disappeared and the line parametrized by $g_1^*$ is therefore unstable in the $g_2$ direction.\\

$\bullet$ The case  $M=2,N>2$ can be studied in a similar way. We find the following fixed point
\beq
g_1^*=?~;~g_2^*=\left\{\begin{array}{c} {-\eps\over N-2}+{\eps^2\over (N-2)^2}\\0\end{array}~;~g_{12}^*=0~~\right.
\eeq
The value $~g_2^*={-\eps\over N-2}+{\eps^2\over (N-2)^2}$ corresponds to the fixed point of the  $N$ colored Potts model \cite{dots2}. We can show that the latter is stable for $g_2^0<0$ otherwise we have a strong coupling regime indicating probably mass generation. We recover the Askhin-Teller fixed line which is stable for $g_1<0$ independently of $g_{12}$.\\

$\bullet$ Finally, if we consider the case $M>2$. We find the fixed points $g_1^*=g_{12}^*=0$. The value of $g_2^*$ depends on the value of N we consider.\\ 
When we consider only one Potts model, we find that $g_{12}^*=0$ is unstable whereas $g_1^*=0$ is stable for $g_1^0<0$. \\
For $N=2$, the only fixed point is the trivial and unstable one. Therefore, the flow is always driven in a strong coupling regime. \\
For $N> 2$, it depends on the initial conditions we consider. For $g_1^0<0$ and $g_2^0<0$ the fixed points $g_1^*=0;~g_2^*={-\eps\over N-2}+{\eps^2\over (N-2)^2};g_{12}^*=0$ is stable. Nevertheless, the situation is uninteresting because the Ising and Potts models are completly decorrelated in the infrared limit. \\

\vskip 0.5cm
We have thus studied in this general section the critical behavior of $M$ Ising models and 
$N$ Potts model coupled each other by their energy density. New non-trivial situations have been exhibited for $M=1$ and $M=2$. We refer to the Figures 1 and 2. The case of one Ising model coupled to several potts models is interesting in so far as it could  offer a possible scenario for the  puzzling physics of the Ising-XY model \cite{isxy}.\\ 
We have also seen
that in many cases, the flow is driven in a strong coupling regime (it often depends on initial conditions). It should be interesting to analyse the effects of weak disorder on such systems in order to test the Imry-Wortis arguments. It will be the matter of the next section.

\section{Coupled minimal models with disorder}
\setcounter{equation}{0}
In this section, we analyse the effects of weak disorder on the previous systems, namely $M$ $q_1$-state Potts models coupled to $N$ $q_2$-state Potts models, with $q_1,q_2\in \{2,3,4\}$. We will pay special attention to the case $q_1=2;~q_2=3$.

\subsection{The model}

The Hamiltonian of the pure system have the following form (see also (\ref{ham}))
\beqa
H&=&\sum\limits_{a=1}^M H_{1}^a+\sum\limits_{b=M+1}^{M+N} H_{2}^b- 
\sum\limits_{c,d} g_{cd} \int d^2 z~\varepsilon^c(z)\varepsilon^d(z)\nn\\&&+m_1\sum\limits_{a=1}^M\int d^2 x \veps_1^a+
m_2\sum\limits_{b=M+1}^{M+N}\int d^2 x \veps_2^b \eeqa
The masses $m_i$ correspond to the reduced temperatures. The addition of a position dependent random coupling constant is equivalent to consider  position dependent random mass terms $m_i\to m_i(x)$ with $\overline{m_i(x)}=m_i$ and  $\overline{m_i(x)m_j(y)}=\Delta_{ij}\delta(x-y)$.
$\Delta_{ij}$ represents the $2\times 2$ symmetric covariance matrix whose 
elements are strictly positive as it should be. If we have considered a diagonal 
covariance matrix corresponding to  independent disorders for each models, then 
$\Delta_{12}$ would have been generated by the R.G. equations. We then apply the 
replicated method by introducing $n$ copies of the system and averaging Gaussian 
distributions for $m_i(x)$. We finally obtain the sum
of $nM$ $q_1$ and $nN$ $q_2$-state Potts models coupled by their energy densities. The
Hamiltonian can be written as
\beq
\label{ham2}
H=H_{1}+H_{2}-\int d^2 x~\sum\limits_{i,j}g_{ij}\sum\limits_{a,b,\al,\be}\veps_i^{a,\al}\veps_j^{b,\al}-\int d^2 
x~\sum\limits_{i,j}\sum\limits_{a,b,\al,\be}\Delta_{ij}^{a,\alpha;b,\beta}\veps_i^{a,\al}\veps_j^{b,\be}~,
\eeq
where $i,j=1,2$, $a,b$ runs from $1$ to $M+N$ depending on the values taken by the lower indices $i,j$  and finally $\al,\be$ run from $1$ to $n$. We consider in the following the replica symmetry scheme. Moreover, a recent work on the random bond Potts model has been performed in order to test the replica symmetry breaking in \cite{rsb}. The results so obtained are in favor of a non replica symmetry breaking senario. The hamiltonian (\ref{ham2}) has therefore six coupling constants $g_1=g_{11},~ g_2=g_{22},~g_{12}=g_{21}$ and $\D_1=\D_{11}^{a,\alpha;b,\beta},~\D_2=\D_{22}^{a,\alpha;b,\beta},~\D_{12}=\D_{21}=\D_{12}^{a,\alpha;b,\beta}$. 
It is therefore a generalisation of the study of $N$ Ising or Potts models with disorder 
\cite{cardy2} \cite{pujol}.

\subsection{Beta functions}
As the model we consider has six coupling constants, we have six beta functions. We have 
shown in appendix D how can we obtain them for $M=N=1$. Th general case goes along the same line but is much more lenghty.  Taking directly the replica limit $n\to 0$, we obtain
\beqa
\label{g1}
\beta_{g_{1}}\equiv {dg_{1}\over d~ln l}&=&\eps_1 g_1 +(M-2)g_1^2-2g_1\D_1+Ng_{12}^2\nn\\&&
+4\D_1^2g_1-(2M-5)\D_1g_1^2-(M-2)g_1^3-Ng_1g_{12}^2-2Ng_1g_{12}\D_{12}\\ \nn\\
\label{d1}
\beta_{\D_{1}}\equiv {d\D_{1}\over d~ln 
l}&=&\eps_1\D_{1}-2\D_1^2+2(M-1)\D_1g_1+2Ng_{12}\D_{12}\nn\\&&
+2\D_1^3-(M-1)g_1\D_1(2\D_1+g_1)-Ng_{12}\D_1(2\D_{12}+g_{12})\\ \nn\\
\label{g2}
\beta_{g_{2}}\equiv {dg_{2}\over d~ln l}&=&\eps_2 g_2 +(N-2)g_2^2-2g_2\D_2+Mg_{12}^2\nn\\&&
+4\D_2^2g_2-(2N-5)\D_2g_2^2-(N-2)g_2^3-Mg_2g_{12}^2-2Mg_2g_{12}\D_{12}\\ \nn\\
\label{d2}
\beta_{\D_{2}}\equiv {d\D_{2}\over d~ln 
l}&=&\eps_2\D_{2}-2\D_2^2+2(N-1)\D_2g_2+2Mg_{12}\D_{12}\nn\\&&
+2\D_2^3-(N-1)g_2\D_2(2\D_2+g_2)-Mg_{12}\D_2(2\D_{12}+g_{12})\\ \nn\\
\label{g12b}
\beta_{g_{12}}\equiv {dg_{12}\over d~ln l}&=&{\eps_1+\eps_2\over 2}g_{12} +(M-1)g_1g_{12}+(N-1)g_2g_{12}-g_{12}(\D_1+\D_2)\nn\\&&
-{1\over 2} (M+N-2) g_{12}^3-{1\over 2}(M-1)g_{12}g_{1}^2-{1\over 2}g_{12}g_2^2+
{1\over 2}(\D_{1}^2+\D_{2}^2)g_{12}\\&&
+3\D_{12}^2g_{12}-(M+N-3)\D_{12}g_{12}^2-(M-1)g_1\D_1g_{12}-(N-1)g_2D_2g_{12}\nonumber\\ \nn\\
\label{d12}
\beta_{\D_{12}}\equiv {d\D_{12}\over d ~ln l}&=&{\eps_1+\eps_2\over 
2}\D_{12}-\D_{12}(\D_{11}+\D_{22})+(M-1)g_1\D_{12}+(N-1)g_2\D_{12}\nn\\&&
+Mg_{12}\D_{1}+Ng_{12}\D_{2}+\D_{12}^3
+{1\over 2}\D_{12}(\D_{1}^2+\D_{2}^2)-{1\over 2}(M-1)g_1\D_{12}(2\D_1+g_1)\nn\\&&-{1\over 2}(N-1))g_2\D_{12}(2\D_2+g_2)
-{1\over 2}(M+N)\D_{12}g_{12}(2\D_{12}+g_{12})
\eeqa
Notice that we recover the results of Pujol \cite{pujol} by taking in these equations the limit $g_{12}=\D_{12}\to 0$. We now want to exploit these equations in order to analyse the fixed point structure in some peculiar cases and in the general one.

\subsection{Study of the fixed point structure}
We first consider some peculiar values of $M$ and $N$. The most simple non trivial case consists in coupling one $q_1$-state to a $q_2$ state Potts model under disorder \cite{simon1}. 

\subsubsection{Case $M=N=1$}
In this case, the beta functions (\ref{g1}-\ref{d12}) reduce to
\beqa
\label{rg11}
\beta_{g_{12}}\equiv {dg_{12}\over d ln l}&=&{\eps_1+\eps_2\over 2}g_{12} - 
g_{12}(\D_{1}+\D_{2})+{1\over 2} 
g_{12}(\D_1^2+\D_2^2)+3\D_{12}^2g_{12}+\D_{12}g_{12}^2\nonumber\\
\beta_{\D_{1}}\equiv {d\D_{1}\over d ln 
l}&=&\eps_1\D_{1}-2\D_{1}^2+2g_{12}\D_{12}+2\D_{1}^3-2\D_{1}\D_{12}g_{12}-\D
_{1}g_{12}^2\nonumber\\
\beta_{\D_{2}}\equiv {d\D_{2}\over d ln l}&=&\eps_2\D_{2}- 2\D_{2}^2+2g_{12}\D_{12}+2\D_{2}^3-2\D_{2}\D_{12}g_{12}-\D_{2}g_{12}^2\\
\beta_{\D_{12}}\equiv {d\D_{12}\over d ln l}&=&{\eps_1+\eps_2\over 
2}\D_{12}+g_{12}(\D_{1}+\D_{2})-\D_{12}(\D_{1}+\D_{2})\nonumber\\
&&+\D_{12}\left[{1\over 
2}(\D_{1}^2+\D_{2}^2)+\D_{12}^2-2\D_{12}g_{12}-g_{12}^2\right]\nonumber
\eeqa
 When we consider one Ising model coupled to a $3$ or $4$-state Potts model, we 
take the limit $\eps_1\to 0$ and  $\eps_2=\eps$, we find the following fixed 
points
\beqa
\label{pfix1}
g^*_{12}=0&;&\D^*_{1}=\D^*_{2}=\D^*_{12}=0\\
\label{pfix2}
g^*_{12}=0;~\D^*_{1}=0;~\D^*_{2}&=&{\eps\over 2}+{\eps^2\over 
4}+o(\eps^2);~\D^*_{12}=0+o(\eps)\\
\label{pfix3}
g^*_{12}=0;~\D^*_{1}=0;~\D^*_{2}&=&{\eps\over 2}+{\eps^2\over 
4}+o(\eps^2);~\D^*_{12}={\sqrt{2}\eps\over 4}+o(\eps)
\eeqa

The first one is trivial and unstable. The second one corresponds to a perfect decoupling of the disordered Ising and Potts models 
(\ref{pfix2}) and we obtain a new fixed point
mixing {\it a priori} both models. Let us notice that $\D^*_{12}$ is 
undetermined at one and two loop calculations just enable to compute the 
first order in $\eps$.  When one considers a $3-$state coupled to a $4-$ state Potts model, 
we only find the fixed points (\ref{pfix1}), (\ref{pfix2}).\\
In order to study the stability of each of these fixed points, we re-express the 
systems  (\ref{rg11}) around the above solutions.  Therefore, we write 
$g_{12}=g^*_{12}+\delta g_{12};~\D_{i}=\D_{i}^*+\delta \D_{i}$ and keep only 
the smallest order in $\eps$. We thus obtain a linear system $\delta \dot{X}=AX$ 
with $X=(\delta\D_{1},\delta\D_{2},\delta\D_{12},\delta g_{12})$. All the 
information concerning the stability is contained in the  matrix $A$. We can  thus see that 
(\ref{pfix1}) and  (\ref{pfix3}) are unstable,  and that
(\ref{pfix2}) is stable.  We have 
represented in Figure 3, the projection on the flow in the 
$(g_{12},\Delta_{12})$ plane for two different initial conditions (points $A_0$ 
and $B_0$). We clearly see that the flow first try to
go away and then is driven by disorder at the origin corresponding to a perfect 
decoupling of the models. As it has been already noticed in \cite{cardy2}, such a 
flow is unusual because it violates the $c-$theorem \cite{zamolod}. This example
 constitutes a new example of first order transition driven by 
randomness 
in a second order one (see Figure 3) and moreover, the models factorize.

Another interesting case concerns the limit $\eps_1=\eps_2=\eps$ which corresponds to two Potts models. We have therefore $g_{12}=g$,
$\Delta_{1}=\Delta_{12}=\Delta_{2}=\Delta$ and the flow reduces to
\beqa
\label{rg2}
{dg \over d\log l} &=& \epsilon g - 2g\Delta + g^2 \Delta
+ 4g \Delta^2 \nonumber\\
{d\Delta \over d\log l} &=& \epsilon \Delta -2 \Delta^2 + 2 \Delta g +2 \Delta^3
-  g^2 \Delta - 2\Delta^2 g
\eeqa
We recover the equations introduced by Pujol \cite{pujol}.
 There are three fixed points
\beq
\label{qfixed1}
g=0~~~;~~~\Delta =0
\eeq
the trivial one, and
\beq
\label{qfixed3}
g=0~~~;~~~\Delta = {\epsilon \over 2} + {\epsilon^2 \over 4}
+ O(\epsilon^3)
\eeq
the fixed point corresponding to a decoupling of the models and
\beq
\label{qfixed4}
g={\epsilon^2 \over 4} + O(\epsilon^3)~~~;
~~~\Delta = {\eps+\eps^2\over 2}+ O(\epsilon^3)
\eeq
a new non-trivial one mixing both models.
The study of the stability of the fixed point
 (\ref{qfixed4}) is non-trivial.
For the initial conditions $g(0)<0$, $\D(0)>0$, the flow is always driven in a strong coupling regime. This is in contradiction with the Imry-Wortis argument \cite{imry}. For $g(0)>0$, $\D(0)>0$, a naive numerical study  shows apparently that the flow is also driven in a strong coupling regime. Nevertheless, we must not forget that the beta functions and therefore the fixed points have been computed order by order in $\eps$ and the numerical and analytical  fixed points can not correspond at all in a few cases. This is exactly what we see in this case. Therefore, we have to take care with a direct interpretation of the numerical flow. Moreover, an analytical study of the fixed point  (\ref{qfixed4}) does not enable to conclude directly since the coefficient $A_{11}$ of the matrix 
of stability $A$ is  zero up to $\eps^2$. A solution could be to include higher order terms and to determine the fixed points up to $\eps^3$. Obviously, the calculations are probably unfeasable. Nevertheless by assuming the general form of the fixed point (\ref{qfixed4}) up to $\eps^3$, we can show that $a_{11}$ does not depend on this form and equals at lowest order in $\eps$ ${\eps^3\over 8}$. By injecting this value in the matrix of stability, we see that the fixed point (\ref{qfixed4}) is in fact stable for $g(0)>0$. It proves in fact that fourth order terms (three loops one) have at least to be included in a numerical analysis of the flow.

\subsubsection{case $~N>1$}

We consider in this section the case of $M$ Ising models coupled to $N$ $q$-state Potts models with disorder. We are looking for the fixed points of the system (\ref{g1}-\ref{d12}). 

First of all, when $M=1$ and $\D_1^*=\D_2^*=\D_{12}^*=0$, we recover the case studied in the section 2.  The value $M=2$ is peculiar since it corresponds to the Ashkin-Teller model. Secondly, by considering only two Ising models, the beta functions associated to $g_1,\D_1$ are those of ref. \cite{vdots}. In this case, the flow can be exactly integrated. When adding the other coupling constants, the situation gets quite more complicated. We have done directly a simulation of the flow in this case. We also find that disorder makes the models decouple as in the previous subsection.

 Otherwise, for general $M\ne 2$, the fixed points are, up to second order in $\eps$, the following (the trivial fixed point is still present)
\beqa
\label{fix1}
&&g_1^*=0~,~\D_1^*=0~;\nn\\
&&g_2^*={-\eps\over (N-2)}+{\eps^2\over (N-2)^2}~,~\D_2^*=0~~{\rm if}~ N>2;\\
&&g_{12}^*=0~,~\D_{12}^*= 0~\nn\\
&&\nn\\
\label{fix2}
&&g_1^*=0~,~\D_1^*=0~;\nn\\
&&g_2^*=0~,~\D_2^*={\eps\over 2}+{\eps^2\over 4};\\
&&g_{12}^*=0~,~\D_{12}^*= 0~{\rm or}~ {\sqrt{2}\over 4}\eps\nn\\
&&\nn\\
\label{fix3}
&&g_1^*=0~,~\D_1^*=0~;\nn\\
&&g_2^*={\eps^2\over 2N}~,~\D_2^*={\eps\over 2}+{(3N-2)\eps^2\over 4N};\\
&&g_{12}^*=0~,~\D_{12}^*= 0~{\rm or}~ {\sqrt{2}\over 4}\eps\nn
\eeqa
The three different fixed points we have found correspond in fact to the three non-trivial fixed points of the ``$N$ colored Potts model with disorder \cite{pujol}. As in the previous subsection, we recover for the fixed points (\ref{fix2}\ref{fix3}) a non trivial value of the disorder mixing {\it a priori}
both models, namely $\D_{12}^*={\sqrt{2}\over 4}\eps$.  Nevertheless, it is straightforward to show that this value is unstable. We can also prove that the fixed point (\ref{fix2}) is unstable. Therefore, only the fixed points (\ref{fix1}) and (\ref{fix3}) (with $\D_{12}^*=0$) are stable.
The flow reaches them depending on initial conditions (see the discussion in \cite{pujol}).
If $N\ne 2$ and $g_2(0)<0$, the flow is always driven in a strong coupling regime as it was already mentioned in the previous subsection (in fact the fixed point (\ref{fix1}) goes to infinity).

Consequently, the infrared physics of $M$ Ising models coupled to $N$ $q-$state Potts models under weak disorder corresponds to $M$ disordered Ising models plus $N$ disordered Potts models. These results can be easily extended to $M$ $3-$state Potts models coupled to $N$ $4-$state Potts models. We also find that disorder makes the models decouple.

\section{Conclusions and Discussions}
In this article, we have studied the critical behavior of different coupled minimal models under weak disorder. But already in the pure case, we obtain some interesting features.

By coupling one Ising model to a several Potts models, we get a non-trivial flow of renormalisation in a three parameter space (see Figure 1). We have found new tricritical points that should correspond to {\it a priori} new conformal theories. Moreover, these results prove that an energy-energy coupling term is a relevant perturbation which can influence the critical behavior. Such a situation might also be encountered in the problem of the XY-Ising model. Indeed, a tricritical point close to the fixed point corresponding to a decoupling of both models can strongly alter numerical results based notably on finite size scaling. It could explain why continuoulsly varying critical exponents are found numerically \cite{isxy}. Moreover, the study of the pure model is also interesting in order to better understand the behavior of several coupled conformal field theories. This approach is complementary to the technics of exact integrability developped recently in \cite{vays,leclair}.

We have also seen that the perturbation term can also drive the system in a strong coupling regime (it depends on initial conditions) indicating probably a first order transition and mass generation. For example, when one considers two different $q$-state Potts models, the flow is always driven in a strong coupling regime. By adding weak randomness, we have found that the first order transitions are often transformed in second order one (see Figure 3) as it should be according to Imry-Wortis argument. Nevertheless, when we consider two coupled Potts models, this argument is violated for some initial conditions in the flow of renormalisation. Moreover, by considering the general case of $M$ $q_1$-state Potts models coupled to $N$ $q_2$-state Potts models under weak randomness, we have show that disorder often makes the models decouple (exept the above case violating Imry-Wortis argument).

Recent numerical results of the disordered $q>4$-state Potts model have shown a puzzling property of the central charge of this model namely $c(q=q_1\times q_2)=c(q_1)+c(q_2)$ \cite{picco2,cardy1}. Therefore, we could have expected some relations between the critical behavior of for example the disordered $6-$state Potts models and one Ising model coupled to $3$-state Potts model with disorder since their central charge are similar.
Nevertheless, the magnetic critical exponent $\beta/\nu\sim0.142$ found in \cite{cardy1}, for the disordered $6-$state Potts model appears clearly different from the one of the Ising or disorderd $3-$state Potts model. Hence, it seems that no relations emerge at a perturbative level between two coupled $q_1$ and $q_2$-state Potts model and a disordered $q_1\times q_2$-state Potts model. Nevertheless, for a large coupling between both Potts models, we could expect some relations between the two situations, but it is outside of our perturbative analysis. That is why it would be very interesting to investigate numerically the critical behavior of several $q-$state Potts models with disorder especially in the strong coupling regime to analyse if we could observe a cross-over phenomena between the two regimes.

\vskip 2cm
{\bf Acknowledgements}\\\noindent
I would like to thank Vl. S. Dotsenko for helpful suggestions and stimulating 
discussions. I also acknowledge M. Picco and P. Pujol for useful discussions.

\eject
\appendix
\section{Beta functions  in the pure case}
\setcounter{equation}{0}
In this appendix, we compute the beta function of $M$ Ising models (indiced by $1$) coupled to $N$ Potts models (indiced by $2$). We have three coupling constant $g_1,~g_2,~g_{12}$. Some terms have already been computed in \cite{dots2}, nevertheless, we recall some details for completeness.

\subsection{First order}
A first  order correction to $g_1$ comes from the contraction of two $
\left(\varepsilon_1 \varepsilon_1 \right) $ terms (plus some combinatorial
factor corresponding to all of the possible contractions) :
$$
{(g_1^0)^2\over2} \int\limits_{|x-y|<r} \displaystyle\sum\limits_{a\not= b=1}^M
\varepsilon_1^a(x) \varepsilon_1^b(x)  \displaystyle\sum_{c\not= d}
\varepsilon_1^c(y) \varepsilon_1^d(y)~ d^2x d^2y
$$
When $b=c\not= a, d$ this contractions  gives~:
$$
 A_1^1(r,\epsilon_1) (g_1^0)^2 \int \displaystyle\sum_{a\not= d}
\varepsilon_1^a(x) \varepsilon_1^d(x) d^2x
$$
Here, $A_1^1(r,\epsilon_1)$  (the lower indice corresponds to the first contribution while the upper indice indicates first order) is the result of the integral produced by the
contraction of two $\varepsilon_1$ operators~:
$$
 A_1^1(r,\epsilon_1) = 2(M-2) \int\limits_{|y-x|<r}
<\varepsilon_1(x)\varepsilon_1(y)>_0 d^2y = 4\pi (M-2)
\int\limits_{y<r}\left(dy\over y^{1+3\epsilon_1}\right)
$$
\beq
 = -4\pi (M-2)\left(r^{-3\epsilon_1}\over 3\epsilon_1\right)
\eeq
Another first order correction to $g_1$ comes from the contraction of two $
\left(\varepsilon_2 \varepsilon_2 \right) $ terms:
$$
{(g_{12}^0)^2\over2} \int\limits_{|x-y|<r} \displaystyle\sum\limits_{a=1}^M\sum \limits_{b=M+1}^{M+N}
\varepsilon_1^a(x) \varepsilon_2^b(x)  \displaystyle\sum_{c=1}^M\sum \limits_{d=M+1}^{M+N}
\varepsilon_1^c(y) \varepsilon_2^d(y) d^2x d^2y
$$
When $b=d$ this contractions  gives~:
$$
 A_2^1(r,\epsilon_2) (g_{12}^0)^2 \int \displaystyle\sum_{a\not= c}
\varepsilon_1^a(x) \varepsilon_1^c(x) d^2x
$$
with
$$
 A_2^1(r,\epsilon_2) = 2N \int\limits_{|y-x|<r}
<\varepsilon_2(x)\varepsilon_2(y)>_0 d^2y = 4\pi N
\int\limits_{y<r}\left(dy\over y^{1+3\epsilon_2}\right)
$$
\beq
 = -4\pi N\left(r^{-3\epsilon_2}\over 3\epsilon_2\right)
\eeq
We obtain similar correction to $g_2$ by changing $(1,M)\leftrightarrow(2,N)$.\\
There are two contributions to $g_{12}$ coming from the contraction of two $(\veps_1\veps_1)$ or  two $(\veps_2\veps_2)$. The integral involved are therefore the same as above and we find

$$
\left( A_3^1(r,\epsilon_1) g_{12}^0g_1^0+  A_4^1(r,\epsilon_2) g_{12}^0g_2^0\right)
\int \displaystyle\sum_{a, c}
\varepsilon_1^a(x) \varepsilon_2^c(x) d^2x
$$
with
\beqa
 A_3^1(r,\epsilon_1) &=& 2(M-1) \left( {r^{-3\epsilon_1}\over 3\epsilon_1} \right)\\
A_4^1(r,\epsilon_2) &=& 2(N-1) \left({r^{-3\epsilon_2}\over 3\epsilon_2}\right)
\eeqa

\subsection{Second  order}
The second order corrections to $g_i$ comes from the contaction of three ($\veps\veps$) terms. We first detail the  second order terms for $g_1$.

\vskip 0.5cm

{\bf Contribution to $g_1$}

\vskip 0.5cm
$\bullet$ The first contribution comes from the contraction of $2\times 2$ $\veps_1$ operators:
\beq
\label{2nn}
{(g_1^0)^3\over3!} \int\limits_{|x-y|,|x-z|<r} \displaystyle\sum\limits_{a\not= b=1}^M
\varepsilon_1^a(x) \varepsilon_1^b(x)  \displaystyle\sum_{c\not= d}
\varepsilon_1^c(y) \varepsilon_1^d(y) \displaystyle\sum_{e\not= f}
\varepsilon_1^e(z) \varepsilon_1^f(z)~ d^2x d^2y d^2z
\eeq
The first one corresponds to the case when $ b = c ; d = e ; b,d \not=
a,f$. We find:
$$
 A_1^2(r,\epsilon_1) (g_1^0)^3 \int \displaystyle\sum_{a\not= f}
\varepsilon_1^a(x) \varepsilon_1^f(x) d^2x
$$
where $A_1^2(r,\epsilon_1)$ corresponds to the following~:
\beq
\label{ee}
 A_1^2(r,\epsilon_1) = 4(M-2)(M-3) \int\limits_{|y-x|,|z-x|<r}
<\varepsilon_1(x)\varepsilon_1(y)>_0 <\varepsilon_1(y)\varepsilon_1(z)>_0 d^2z d^2y
\eeq
We replace  $<\varepsilon(x)_1\varepsilon(y)_1>_0 $ by
$|x-y|^{-2-3\epsilon_1}$. This integral has been performed in \cite{dots2} and we have
\beqa
\label{atu}
A_1^2(r,\epsilon_2) &=&  8\pi (M-2)(M-3)
\int\limits_{y<r}\left(dy\over y^{1+6\epsilon_1}\right) \int
|z|^{-2-3\epsilon_1} |z-1|^{-2-3\epsilon_1} d^2 z \nn\\
&= & 16\pi^2 (M-2)(M-3)\left(r^{-6\epsilon_1}\over
9\epsilon_1^2\right)
\eeqa
The second term is produced when $a = c = e \not= f; b = d \not= f$. This
will be denoted by~:
$$
A_2^2(r,\epsilon_1)(g_1^0)^3 \int \displaystyle\sum_{a\not= f}
\varepsilon_1^a(x) \varepsilon_1^f(x) d^2x
$$
with:
\beq
\label{gg}
 A_2^2(r,\epsilon_1) = 4 (M-2)  \int\limits_{|y-x|,|z-x|<r}
<\varepsilon_1(x) \varepsilon_1(y) \varepsilon_1(z) \varepsilon_1(\infty)>_0 <
\varepsilon_1(y)  \varepsilon_1(z)>_0 d^2y d^2z
\eeq
The calculation of this integral has been made in \cite{dots2}. We will nevertheless give the main steps because we will have to compute other integrals using similar technics.\\
 The $<\varepsilon_1(x) \varepsilon_1(y) \varepsilon_1(z)
\varepsilon_1(\infty)>_0$ term corresponds to the result of $ \varepsilon_1
\varepsilon_1 \varepsilon_1 \rightarrow \varepsilon_1$ obtained by projecting
$\varepsilon_1 \varepsilon_1 \varepsilon_1 $ over $\varepsilon_1 (\infty)$. We know use the Coulomb gas formulation
\beqa
A_2^2(r,\epsilon_1) &=& 4 (M-2) {\cal N} \times \nn \\
 \int\limits_{|y-x|,|z-x|<r}& & \!\!\!\!\!\!\!\!\!\!\!\!\!\!\!
<V_{\overline{12}}(x) V_{12}(y) V_{12}(z) V_+(u) V_{12}(\infty)>
|y-z|^{-2-3\epsilon_1} d^2y d^2z d^2u,
\eeqa

\no where ${\cal N}$ is a normalisation constant which is fixed by taking the limit $R\to 0$ in the four points function. Indeed, we thus have 
$$<\varepsilon_1(0) \varepsilon_1(R) \varepsilon(x) \varepsilon(y)> =  {\cal N} \int
<V_{\overline{12}}(0) V_{12}(R) V_{12}(x) V_+(u) V_{12}(y)> d^2u~.
$$
By taking the limit $R\to 0$ in both sides, we therefore obtain 
\beq
{\cal N}= -{2\over \sqrt{3}} {(\Gamma(-{2 \over 3}))^2 \over(\Gamma(-{1 \over 3}))^4}~.
\eeq
In order to compute (\ref{gg}), we first redefine $
y \rightarrow y-x; z \rightarrow z-x$. Then, the idea is to pass to the ``gauge''
$z_1=0,~z_2=z,~z_3=1, z_4\to\infty$. For arbitrary value of $z_i;~i\in[1..4]$, we can show using different changes of variables that
\beqa
\label{cdv}
I&=&\int d^2 u <V_{\alpha_1}(z_1)V_{\alpha_2}(z_2)V_{\alpha_3}(z_3)V_{\alpha_4}(z_4) V_+(u)>
\nn\\&&\nn\\&\propto & |z_{12}|^{4\alpha_1\alpha_2}|z_{13}|^{4\alpha_1\alpha_3}|z_{23}|^{4\alpha_2\alpha_3}\nn\\
&&|z_{14}|^{2+4(\alpha_1+\alpha_4)\alpha_+ +4\alpha_1\alpha_4}
|z_{24}|^{4\alpha_2\alpha_++4\alpha_2\alpha_4}
|z_{34}|^{2+4(\alpha_3+\alpha_4)\alpha_++4\alpha_3\alpha_4} \nn\\
 &&\int d^2 u |u|^{4\alpha_1\alpha_+}|u-\eta|^{4\alpha_2\alpha_+}|u-1|^{4\alpha_3\alpha_+},
\eeqa
with $\alpha_1+\alpha_2+\alpha_3+\alpha_4+\alpha_+=2\alpha_0$ as it should be, and $\eta={z_{12}z_{34}\over z_{13}z_{14}}$. Using the above formula (\ref{cdv}) and two trivial changes of variables, we find
\beqa
\label{ia22}
A_2^2(r,\epsilon_1) &=& 8 \pi (M-2) {\cal N} \times \\
&&\int\limits_{y<r}\left(dy\over y^{1+6\epsilon}\right)
\int |z|^{-4\Delta_{12}}  |z-1|^{-4\Delta_{12}+4\alpha^2_{12}}
|u|^{4\alpha_+ \alpha_{\overline{12}}} |u-1|^{4\alpha_+ \alpha_{12}}
|u-z|^{4\alpha_+ \alpha_{12}} d^2z d^2u \nn
\eeqa
The second integral is computed in \cite{dots2} and recalled in appendix E (eq: (\ref{res1}). It equals $\sqrt{3}\pi{\Gamma^4(-{1\over3})\over
\Gamma^2(-{2\over 3})} $. In the $z$ integration, we introduce a new singularity at infinity which need to be substracted (see \cite{dots2} for more details). We finally obtain
\beq 
A_2^2(r,\epsilon_1)=8\pi^2(M-2) \left(r^{-6\epsilon_1}\over
3\epsilon_1\right)\left[1+{2\over 3\eps_1}\right]
\eeq

$\bullet$The second type of terms comes from two $\veps_1$  and two $\veps_2$ contractions.
namely
\beq
{g_1^0(g_{12}^0)^2\over3!} \int\limits_{|x-y|,|x-z|<r} \displaystyle\sum\limits_{a\not= b=1}^M
\varepsilon_1^a(x) \varepsilon_1^b(x)  \displaystyle\sum_{c,d}
\varepsilon_1^c(y) \varepsilon_2^d(y) \displaystyle\sum_{e,f}
\varepsilon_1^e(z) \varepsilon_2^f(z) d^2x d^2y d^2z
\eeq
The first contribution comes from the case $b=c\ne a,e;~d=f$. We denote this term by 
$$A_3^2(r,\epsilon_1,\eps_2) g_1^0(g_{12}^0)^2\int\sum\limits_{a\ne e}\varepsilon_1^a(x)
\varepsilon_1^e(x) d^2x$$ with
\beq 
\label{a32}
A_3^2(r,\epsilon_1,\eps_2)=8N(M-2) \int\limits_{|y-x|,|z-x|<r}
<\varepsilon_1(x)\varepsilon_1(y)>_0 <\varepsilon_2(y)\varepsilon_2(z)>_0 d^2z d^2y
\eeq
Using standard manipulations, we find
\beq
A_3^2(r,\epsilon_1,\eps_2)=32\pi^2 N (M-2) {r^{-3(\eps_1+\eps_2)}\over 9\eps_1\eps_2}
\eeq

The second contribution comes from the case $a=c=e; d=f$ and is denoted by
$$A_4^2(r,\epsilon_1,\eps_2)g_1^0(g_{12}^0)^2\int\sum\limits_{a\ne e}\varepsilon_1^a(x)
\varepsilon_1^e(x) d^2x$$
with
\beq
 A_4^2(r,\epsilon_1,\eps_2) = 4 N  \int\limits_{|y-x|,|z-x|<r}
<\varepsilon_1(x) \varepsilon_1(y) \varepsilon_1(z) \varepsilon_1(\infty)>_0 <
\varepsilon_2(y)  \varepsilon_2(z)>_0 d^2y d^2z
\eeq
Using the same transformations explained in the calculation of $A_2^2$, we find
\beqa
\label{ia42}
A_4^2(r,\epsilon_1,\eps_2) &=& 8 \pi N ~{\cal N} \times \\
&&\int\limits_{y<r}\left(dy\over y^{1+3\epsilon_1+3\eps_2}\right)
\int |z|^{-4\Delta_{12}}  |z-1|^{-4\Delta'_{12}+4\alpha^2_{12}}
|u|^{4\alpha_+ \alpha_{\overline{12}}} |u-1|^{4\alpha_+ \alpha_{12}}
|u-z|^{4\alpha_+ \alpha_{12}} d^2z d^2u \nn
\eeqa
with $4\Delta'_{12}=-2-3\eps_2$ while $4\Delta_{12}=-2-3\eps_1$.
The first integral equals $-{r^{-3(\eps_1+\eps_2)}\over3(\eps_1+\eps_2)}$. The second integral is computed in the appendix E (eq:(\ref{res2})) and equals
\beq
\pi\sqrt{3}{\Gamma^4(-{1\over3})\over\Gamma^2(-{2\over 3})}\left[1+{1\over 3\eps_1}-{1\over 3\eps_2}\right]
\eeq
As for the calculation of $A_2^2$, the $z$ integration in (\ref{ia42}) is performed over the whole complex plane, while the domain of integration is in fact restricted to the disk $|z|<{r\over |y|}$. This introduces a new singularity at infinity. This is equivalent for this integral to add the singularity at the origin (see \cite{dots2}). Some easy manipulations show that this extra contribution from $z\to 0$ introduces an integral similar as in (\ref{a32}).
We finally obtain
\beqa
\label{a42}
A_4^2(r,\epsilon_1,\eps_2)&=&16\pi^2 {r^{-3(\eps_1+\eps_2)}\over3(\eps_1+\eps_2)}\left[1+{1\over 3\eps_1}-{1\over 3\eps_2}\right]+16\pi^2 {r^{-3(\eps_1+\eps_2)}\over 9\eps_1\eps_2}\nn\\
&=&16\pi^2 {r^{-3(\eps_1+\eps_2)}\over3(\eps_1+\eps_2)}\left[1+{2\over 3\eps_1}\right]
\eeqa
$\bullet$ There is finally a last contribution to $g_1$ corresponding to the contraction of $2\times 2$ $\veps_2$ operators. It comes from
$$
g_2^0 (g_{12}^0)^2  \int\limits_{|x-y|,|x-z|<r} \displaystyle\sum\limits_{a\not= b=M+1}^{M+N}
\varepsilon_2^a(x) \varepsilon_2^b(x)  \displaystyle\sum_{c,d}
\varepsilon_1^c(y) \varepsilon_2^d(y) \displaystyle\sum_{e,f}
\varepsilon_1^e(z) \varepsilon_2^f(z) d^2x d^2y d^2z
$$
The contribution comes from the case $a=d;b=f$. We denote this case by
$$
g_2^0 (g_{12}^0)^2 A_5^2(r,\epsilon_1,\eps_2)  \int \sum\limits_{c\not=e}\veps_1^c\veps_1^e$$
The computation of $A_5^2(r,\epsilon_1,\eps_2)$ is straighforward and we find
\beq
\label{a52}
A_5^2(r,\epsilon_1,\eps_2)=16\pi^2 N(N-1){r^{-6\eps_2}\over 9\eps_2^2}
\eeq
 When we gather all these contributions to $g_1$, we find the equations (\ref{renor1}).
Obviously, by exchanging $(1,M)\leftrightarrow (2,N)$, we obtain the whole contribution to $g_2$ (see eq. (\ref{renor2})).

\vskip 0.5cm

{\bf Contribution to $g_{12}$}

\vskip 0.5cm

$\bullet$
The first contribution to $g_{12}$ happens with contractions of $\veps_1$ operators. It comes from
$$
(g_1^0)^2 g_{12}^0 \int\limits_{|x-y|,|x-z|<r} \displaystyle\sum\limits_{a\not= b=1}^{M}
\varepsilon_1^a(x) \varepsilon_1^b(x)  \displaystyle\sum_{c\ne d}
\varepsilon_1^c(y) \varepsilon_1^d(y) \displaystyle\sum_{e,f}
\varepsilon_1^e(z) \varepsilon_2^f(z) d^2x d^2y d^2z
$$
We then have two possible contractions.
We write the result in the form 
$$
g_{12}^0 (g_{1}^0)^2 A_6^2(r,\epsilon_1,\eps_2)  \int \sum\limits_{a,b}\veps_1^a\veps_2^b$$

We then follow  calculations  similar to $A_1^2$ and $A_2^2$ (except the symmetry factors) and find
\beq
\label{a62}
A_6^2(r,\epsilon_1,\eps_2)=4\pi^2 (M-1){r^{-6\eps_1}\over 3\eps_1}\left[1+{2(M-1)\over 3\eps_1}\right]
\eeq
We have also the symmetric contribution coming from $\veps_2$ contractions, namely
\beq
\label{a72}
A_7^2(r,\epsilon_1,\eps_2)=4\pi^2 (N-1){r^{-6\eps_2}\over 3\eps_2}\left[1+{2(N-1)\over 3\eps_2}\right]
\eeq

$\bullet$
We then have the contribution coming from the contraction of two $\veps_1$ and two $\veps_2$
$$
g_1^0g_2^0 g_{12}^0 \int\limits_{|x-y|,|x-z|<r} \displaystyle\sum\limits_{a\not= b=1}^{M}
\varepsilon_1^a(x) \varepsilon_1^b(x)  \displaystyle\sum\limits_{c\not= d=M+1}^{M+N}
\varepsilon_2^c(y) \varepsilon_2^d(y) \displaystyle\sum_{e,f}
\varepsilon_1^e(z) \varepsilon_2^f(z) d^2x d^2y d^2z
$$
By taking $a=e\ne b;c=f \ne d$, we obtain 
$$
g_1^0g_2^0 g_{12}^0  A_8^2(r,\epsilon_1,\eps_2)\int \sum\limits_{a,b}\veps_1^a\veps_2^b$$
with 
\beq
\label{a82}
A_8^2(r,\epsilon_1,\eps_2)=16\pi^2 (N-1)(M-1){r^{-3(\eps_1+\eps_2)}\over 9\eps_1\eps_2}
\eeq
\no
$\bullet$
Finally, we have terms coming from $(g_{12}^0)^3$ (written as $(g_{12}^0)^3 A_9^2(r,\epsilon_1,\eps_2)\int \sum\limits_{a,b}\veps_1^a\veps_2^b$)
$$
(g_{12}^0)^3 \int\limits_{|x-y|,|x-z|<r} \displaystyle\sum\limits_{a, b}
\varepsilon_1^a(x) \varepsilon_2^b(x)  \displaystyle\sum\limits_{c,d}
\varepsilon_1^c(y) \varepsilon_2^d(y) \displaystyle\sum_{e,f}
\varepsilon_1^e(z) \varepsilon_2^f(z) d^2x d^2y d^2z
$$
We have three different possibilities: first contracting two $\veps_1$ and two $\veps_2$, 
second contracting three $\veps_1$ (by the operator algebra $\veps_1\veps_1\veps_1\to \veps_1$) and two $\veps_2$ and third contracting  three $\veps_2$ and two $\veps_1$.
The first one leads to a similar contribution as $A_8^2$. The second and third case involve similar integrals as for $A_4^2$. We find
\beqa
\label{a92}
A_9^2(r,\epsilon_1,\eps_2)&=&16\pi^2 (N-1)(M-1){r^{-3(\eps_1+\eps_2)}\over 9\eps_1\eps_2}\nn\\
&&+8\pi^2{r^{-3(\eps_1+\eps_2)}\over 3(\eps_1+\eps_2)}\left[M+N-2+{2(M-1)\over 3\eps_1}+ {2(N-1)\over 3\eps_2}\right]
\eeqa
By gathering all these contributions, we recover the equation (\ref{renor3}). 

\section{Renormalisation of $m_i$ in the pure case}
\setcounter{equation}{0}
This appendix is devoted to to the computation  of the renormalisation of the
coupling constants $m_i$ associated to the energy operator $\veps_i$. As in the appendix A,
only contraction between different $\veps_i$ are involved.

\subsection{First order}
There are two contributions at this order to $m_1$. The first one comes from the term
$$
g_1^0 m_1^0 \int\limits_{|x-y|<r} \displaystyle\sum\limits_{a\not= b=1}^M
\varepsilon_1^a(x) \varepsilon_1^b(x)  \displaystyle\sum\limits_{c=1}^M
\varepsilon_1^c(y) d^2x d^2y
$$
When  $ b=c$, we can contract two $\varepsilon_1 $ operators~:
$$
g_1^0 m_1^0 B_1(r,\epsilon_1) \int\displaystyle\sum_{a} \varepsilon_1^a(y) d^2y
$$
with
\beq
 B_1^1(r,\epsilon_1) = 2(M-1) \int\limits_{|y-x|<r}
<\varepsilon_1(x)\varepsilon_1(y)>_0 d^2y = -4\pi
(M-1)\left(r^{-3\epsilon_1}\over
3\epsilon_1\right)
\eeq

The second contribution comes from the contraction of an $(g_{12}^0)\veps_1\veps_2$ with
$m_2^0 \veps_2$
$$
g_{12}^0 m_2^0 \int\limits_{|x-y|<r} \displaystyle\sum\limits_{a,b}
\varepsilon_1^a(x) \varepsilon_2^b(x)  \displaystyle\sum\limits_{c=M+1}^{M+N}
\varepsilon_2^c(y) d^2x d^2y
$$
When  $ b=c$, we can contract two $\varepsilon_2 $ operators~:
$$
g_{12}^0 m_2^0 B_2^1(r,\epsilon_1,\eps_2) \int\displaystyle\sum_{a} \varepsilon_1^a(y) d^2y
$$
with
\beq
 B_2^1(r,\epsilon_1,\eps_2) = 2N \int\limits_{|y-x|<r}
<\varepsilon_2(x)\varepsilon_2(y)>_0 d^2y = -4\pi
N\left(r^{-3\epsilon_2}\over
3\epsilon_2\right)
\eeq
This is precisely this term which mix both masses. Obviously, we have by symmetry similar contibutions for $m_2$.

\subsection{Second  order}
We now compute the contributions at second order to $m_1$. The first involves only 
$\veps_1$ operators
$$
{(g_1^0)^2\over2} m_1^0 \int\limits_{|x-y|,|x-z|<r} \displaystyle\sum\limits_{a\not= b=1}^M
\varepsilon_1^a(x) \varepsilon_1^b(x)  \displaystyle\sum_{c\not= d}
\varepsilon_1^c(y) \varepsilon_1^d(y) \displaystyle\sum_{e}
\varepsilon_1^e(z)  d^2x d^2y d^2z
$$
This expression is very similar to the one involved in the calculations of $A_1^2$ and $A_2^2$. Using the same reasoning, we find
$$
(g_1^0)^2 m_1^0 B_1^2(r,\eps_1)\int\displaystyle\sum_{e}\varepsilon_1^a(x)  d^2x
$$
with 
\beqa
\label{b12}
 B_1^2(r,\eps_1)&=&16\pi^2(M-1)(M-2)\left(r^{-6\epsilon_1}\over
9\epsilon_1^2\right)+ 4 \pi^2 (M-1) \left(r^{-6\epsilon_1}\over
3\epsilon_1\right)\left[1+{2\over 3\eps_1}\right]\nn\\
&=& 4 \pi^2 (M-1) \left(r^{-6\epsilon_1}\over
3\epsilon_1\right)\left[1+{4M-6\over 3\eps_1}\right]
\eeqa
The second contribution comes from
$$
{(g_{12}^0)^2\over2} m_1^0 \int\limits_{|x-y|,|x-z|<r} \displaystyle\sum\limits_{a,b}
\varepsilon_1^a(x) \varepsilon_2^b(x)  \displaystyle\sum_{c,d}
\varepsilon_1^c(y) \varepsilon_2^d(y) \displaystyle\sum_{e}
\varepsilon_1^e(z)  d^2x d^2y d^2z
$$
In this case, we then follow the calculations of $A_3^2$ and $A_4^2$. We find 

$$
(g_{12}^0)^2 m_1^0 B_2^2(r,\eps_1,\eps_2)\int\displaystyle\sum_{e}\varepsilon_1^a(x)  d^2x
$$
with 
\beqa
\label{b22}
 B_1^2(r,\eps_1,\eps_2)&=&16\pi^2N(M-1)\left(r^{-3(\epsilon_1+\eps_2)}\over
9\epsilon_1\eps_2\right)+ 8 \pi^2 N \left(r^{-3(\epsilon_1+\eps_2)}\over
3(\epsilon_1+\eps_2)\right)\left[1+{2\over 3\eps_1}\right]\nn\\
&=& 4 \pi^2 N \left(r^{-3(\epsilon_1+\eps_2)}\over
3(\epsilon_1+\eps_2)\right)\left[1+{2M\over 3\eps_1} +{2(M-1)\over 3\eps_2}\right]
\eeqa

We have finally the contributions coming from $m_2^0$ at second order, namely
$$
{g_{12}^0 g_1^0\over2} m_2^0 \int\limits_{|x-y|,|x-z|<r} \displaystyle\sum\limits_{a\ne b=1}^M
\varepsilon_1^a(x) \varepsilon_1^b(x)  \displaystyle\sum_{c,d}
\varepsilon_1^c(y) \varepsilon_2^d(y) \displaystyle\sum\limits_{e+M+1}^{M+N}
\varepsilon_2^e(z)  d^2x d^2y d^2z
$$
and
$$
{g_{12}^0 g_2^0\over2} m_2^0 \int\limits_{|x-y|,|x-z|<r} \displaystyle\sum\limits_{a\ne b=M+1}^{M+N}
\varepsilon_2^a(x) \varepsilon_2^b(x)  \displaystyle\sum_{c,d}
\varepsilon_1^c(y) \varepsilon_2^d(y) \displaystyle\sum\limits_{e+M+1}^{M+N}
\varepsilon_2^e(z)  d^2x d^2y d^2z
$$

Taking $b=c; d=e$ in the first one and $a=e; b=d$ in the second one we obtain
$$\left(g_{12}^0 g_1^0 m_2^0B_3^2(r,\eps_1,\eps_2)+g_{12}^0 g_2^0 m_2^0B_4^2(r,\eps_1,\eps_2\right))\int\displaystyle\sum_{e}\varepsilon_1^a(x)  d^2x
$$
with
\beqa 
B_3^2(r,\eps_1,\eps_2)&=& 8\pi^2N(M-1)\left(r^{-3(\epsilon_1+\eps_2)}\over
9\epsilon_1\eps_2\right)\\
B_4^2(r,\eps_1,\eps_2)&=& 8\pi^2N(N-1)\left(r^{-6\eps_2)}\over
9\epsilon_2^2\right)
\eeqa
By collecting all these contributions, we obtain the equations (\ref{mas1}) and by symmetry the equation (\ref{mas2}).

\section{Renormalisation of $h_i$ in the pure case}
\setcounter{equation}{0}
We now compute the renormalisation of the coupling constants $h_i(r)$ associated to the spin operators $\s_i$ ($\s_1$ the Ising spin operator, $\s_2$ a Potts spin operator).
We restrict ourself to the most interesting case, namely $M=1$ Ising model coupled to $N$ Potts models. Here, the renormalisation of the spin operators is induced by the interaction term (see eq: ()) containing only energy operators. Therefore, we have a mixture of $\s_i$ with $\veps_i$. The first order term gives no contribution because the OPE of two $\veps$ operators does not involve $\s$ operators. The first non-zero contribution is at second order.

\subsection{Second  order}
Here, we compute the renormalisation of a Potts spin operator with the interaction term
$\sum\limits_{c,d}\sum\limits_{i,j} g_{ij} \int d^2 z~\varepsilon_i^c(z)\varepsilon_j^d(z)$. There are two non-zero contributions. The first one comes from
\beq
{(g_1^0)^2\over 2}  h_1^0 \int\limits_{|x-y|,|x-z|<r} \displaystyle\sum_{a\not= b}
\varepsilon_1^a(x) \varepsilon_1^b(x)  \displaystyle\sum_{c\not= d}
\varepsilon_1^c(y) \varepsilon_1^d(y) \displaystyle\sum_{e}
\sigma_1^e(z)  d^2x d^2y d^2z
\eeq
A similar integral has been encountered in \cite{dots2}. When $ a=c=e\not=b=d$ the product  $\sigma_1\varepsilon_1\varepsilon_1$
contains a $\sigma$ operator. As usual, we project the product $\sigma_1\varepsilon_1\varepsilon_1$ on a $\sigma_1(\infty)$ operator, obtaining therefore
$$
 h_1^0  (g_1^0)^2 C_1^{(2)}(r,\epsilon_1,\eps_2)\int\displaystyle\sum_{a=1}^M \sigma_1^a(z) d^2z
$$
where
\beq
 C_1^{(2)}(r,\epsilon_1,\eps_2) = 2 (N-1)  \int\limits_{|y-x|,|z-x|<r}
<\sigma_1(x) \varepsilon_1(y) \varepsilon_1(z) \sigma_1(\infty)>_0 <
\varepsilon_1(y)  \varepsilon_1(z)>_0 d^2y d^2z
\eeq
Using the Coulomb gas formulation, this integral reads
\beqa
\label{ic12}
& & 4 (M-1)~{\cal N} \pi \int\limits_{y<r}\left(dy\over y^{1+6\epsilon_1}\right)
\times\\
& & \int |z|^{4\alpha_{12}\alpha_{\overline{k,k-1}}}
|z-1|^{-4\Delta_{12}+4\alpha^2_{12}}
|u|^{4\alpha_+ \alpha_{\overline{k,k-1}}} |u-1|^{4\alpha_+ \alpha_{12}}
|u-z|^{4\alpha_+ \alpha_{12}} d^2z d^2u \nn
\eeqa
where the normalisation ${\cal N}$ is the same as the one computed in the appendix A. The second non-trivial has been computed in \cite{dots2} (see also the equation (\ref{res3}) for such integrals)
and we find

\beq
 C_1^{(2)}(r,\epsilon_1,\eps_2) = (M-1) \pi^2
{r^{-6\epsilon_1}\over 2}\left[1 + {4\over3}(2-\lambda_1)
{\Gamma^2(-{2\over3})\Gamma^2({1\over6})\over\Gamma^2(-{1\over3})
\Gamma^2(-{1\over6})}\right]
\eeq

The other one comes from
\beq
{(g_{12}^0)^2\over 2}  h_1^0 \int\limits_{|x-y|,|x-z|<r} \displaystyle\sum_{a, b}
\varepsilon_1^a(x) \varepsilon_2^b(x)  \displaystyle\sum_{c, d}
\varepsilon_1^c(y) \varepsilon_2^d(y) \displaystyle\sum_{e}
\sigma_1^e(z)  d^2x d^2y d^2z
\eeq
We take $a=c=e$, $b=d$ and project on the $\s_1(\infty)$ operator and thus obtain
$$
 h_1^0  (g_{12}^0)^2 C_2^{(2)}(r,\epsilon_1,\eps_2)\int\displaystyle\sum_{a=1}^M \sigma_1^a(z) d^2z
$$
with 
\beq
 C_2^{(2)}(r,\epsilon_1,\eps_2) = 2N  \int\limits_{|y-x|,|z-x|<r}
<\sigma_1(x) \varepsilon_1(y) \varepsilon_1(z) \sigma_1(\infty)>_0 <
\varepsilon_2(y)  \varepsilon_2(z)>_0 d^2y d^2z
\eeq
Using the Coulomb gas formalism, we obtain an integral very similar to (\ref{ic12})
\beqa
\label{ic22}
& & 4 N~{\cal N} \pi \int\limits_{y<r}\left(dy\over y^{1+3\epsilon_1+3\eps_2}\right)
\times\\
& & \int |z|^{4\alpha_{12}\alpha_{\overline{k,k-1}}}
|z-1|^{-4\Delta'_{12}+4\alpha^2_{12}}
|u|^{4\alpha_+ \alpha_{\overline{k,k-1}}} |u-1|^{4\alpha_+ \alpha_{12}}
|u-z|^{4\alpha_+ \alpha_{12}} d^2z d^2u \nn
\eeqa
with $4\Delta'_{12}=-2-3\eps_2$ whereas $4\Delta_{12}=-2-3\eps_1$. This integral has been computed in the appendix E (see (\ref{res4})) and yields 

\beq
 C_2^{(2)}(r,\epsilon_1,\eps_2) = {3N\over 2} \pi^2 {r^{-3(\eps_1+\eps_2)}\over 3(\eps_1+\eps_2)}\left[\eps_2+\eps_1(1+ {8\over3}(2-\lambda_1)
{\Gamma^2(-{2\over3})\Gamma^2({1\over6})\over\Gamma^2(-{1\over3})
\Gamma^2(-{1\over6})}\right]
\eeq
Obviously, we get similar contributions for $h_2^0$ when we substitute $(1,M)\leftrightarrow (2,N)$.

\subsection{third  order}
At third order, we have to do the contraction of threee $(\eps\eps)$ operators with one $\s$ operator.
We obtain three main contributions. 

$\bullet$The first one does not mix the $\veps_1$ and $\veps_2$ operators and has been computed in \cite{dots2}. It reads as
$$
{(g_1^0)^3\over3!} h_1^0 \int\limits_{|x-y|,|x-z|,|x-u| <r}
\displaystyle\sum_{a\not= b}
\varepsilon_1^a(x) \varepsilon_1^b(x)  \displaystyle\sum_{c\not= d}
\varepsilon_1^c(y) \varepsilon_1^d(y)
 \displaystyle\sum_{e\not= f}\varepsilon_1^e(x) \varepsilon_1^f(x)
\displaystyle\sum_{g}\sigma_1^g(z)  d^2x d^2y d^2z d^2u
$$
Here contributions  come from the following contractions
$a=c=g ; b=e ; d=f$ and produce~:
$$
(g_1^0)^3 h_1^0  C_1^3(r,\epsilon_1)\int\displaystyle\sum_{a=1}^M \sigma_1^a(z) d^2z
$$
where
\beqa
 C_1^3(r,\epsilon_1) &=& 4(M-1)(M-2) \times \\
 \int\limits_{|y-x|,|z-x|,|w-x| <r}&&\!\!\!\!\!\!\!\!\!\!\!\!\!\!
<\sigma_1(x) \varepsilon_1(y) \varepsilon_1(z) \sigma_1(\infty)>_0
<\varepsilon_1(y)  \varepsilon_1(w)>_0<\varepsilon_1(w)  \varepsilon_1(z)>_0 d^2y
d^2z d^2w \nn
\eeqa
and leads in the Coulomb gas representation to
$$
 8(M-1)(M-2){\cal N} \pi \int
|w|^{-2-3\epsilon_1} |w-1|^{-2-3\epsilon_1} d^2w\int\limits_{y<r}\left (dy\over
y^{1+9\epsilon_1}\right)
$$
\beq
\label{ic13}
\int |z|^{4\alpha_{12}\alpha_{\overline{k,k-1}}}
|z-1|^{-8\Delta_{12}+4\alpha^2_{12}+2}
|u|^{4\alpha_+ \alpha_{\overline{k,k-1}}} |u-1|^{4\alpha_+ \alpha_{12}}
|u-z|^{4\alpha_+ \alpha_{12}} d^2z d^2u~.
\eeq
In order to factorize the $y$ and $w$ integrations, we have used the formula (\ref{cdv}) and three trivial changes of variables.
 The first two
integrations will then produce ${24\pi r^{-9\epsilon_1} \over 81 \epsilon_1^2}
$ while the results of the $z$ and $u$
integrations are given in \cite{dots2} and in   appendix E (\ref{res5}). Therefore, the final result is
\beq
C_1^3(r,\epsilon_1) = -12(M-1)(M-2) \pi^3
\left(r^{-9\epsilon-1}\over9\epsilon-1\right)\left[1 + {8\over9}(2-\lambda_1)
{\Gamma^2(-{2\over3})\Gamma^2({1\over6})\over\Gamma^2(-{1\over3})
\Gamma^2(-{1\over6})}\right]
\eeq

$\bullet$ The second contribution reads as
$$
{(g_{12}^0)^2\over3!}g_1^0 h_1^0 \int\limits_{|x-y|,|x-z|,|x-u| <r}
\displaystyle\sum_{a\not= b}
\varepsilon_1^a(x) \varepsilon_1^b(x)  \displaystyle\sum_{c, d}
\varepsilon_1^c(y) \varepsilon_2^d(y)
 \displaystyle\sum_{e, f}\varepsilon_1^e(x) \varepsilon_2^f(x)
\displaystyle\sum_{g}\sigma_1^g(z)  d^2x d^2y d^2z d^2u
$$
Taking for example
$a=c=g ; b=e ; d=f$, we obtain
$$
(g_{12}^0)^2 g_1^0 C_2^3(r,\epsilon_1,\eps_2)\int\displaystyle\sum_{a=1}^M \sigma_1^a(z) d^2z
$$
where
\beqa
 C_2^3(r,\epsilon_1) &=& 8N(M-1) \times \\
 \int\limits_{|y-x|,|z-x|,|w-x| <r}&&\!\!\!\!\!\!\!\!\!\!\!\!\!\!
<\sigma_1(x) \varepsilon_1(y) \varepsilon_1(z) \sigma_1(\infty)>_0
<\varepsilon_1(y)  \varepsilon_1(w)>_0<\varepsilon_2(w)  \varepsilon_2(z)>_0 d^2y
d^2z d^2w \nn
\eeqa
and leads in the Coulomb gas representation to
$$
 16N(M-1) {\cal N} \pi \int
|w|^{-2-3\epsilon_1} |w-1|^{-2-3\epsilon_2} d^2w\int\limits_{y<r}\left (dy\over
y^{1+6\epsilon_1+3\eps_2}\right)
$$
\beq
\label{ic23}
\int |z|^{4\alpha_{12}\alpha_{\overline{k,k-1}}}
|z-1|^{-4\Delta_{12}-4\Delta'_{12}+4\alpha^2_{12}+2}
|u|^{4\alpha_+ \alpha_{\overline{k,k-1}}} |u-1|^{4\alpha_+ \alpha_{12}}
|u-z|^{4\alpha_+ \alpha_{12}} d^2z d^2u~.
\eeq
The third integral is computed in appendix E (see (\ref{res6})) and the final result is
\beq
C_2^3(r,\epsilon_1,\eps_2) =
-8N(M-1)\pi^3 (g_{12}^0)^2g_1^0 {r^{-6\eps_1-3\eps_2}\over 6\eps_1+3\eps_2 }\left[(1+{\eps_1\over \eps_2})(1+{4\over 3}(2-\lambda_1){\cal R})+{1\over 2}(1+{\eps_2\over \eps_1}) \right].
\eeq
$\bullet$ The third contribution reads as
$$
{(g_{12}^0)^2\over3!}g_2^0 h_1^0 \int\limits_{|x-y|,|x-z|,|x-u| <r}
\displaystyle\sum_{a, b}
\varepsilon_1^a(x) \varepsilon_2^b(x)  \displaystyle\sum_{c, d}
\varepsilon_1^c(y) \varepsilon_2^d(y)
 \displaystyle\sum\limits_{e\ne f=M+1}^{M+N}\varepsilon_2^e(x) \varepsilon_2^f(x)
\displaystyle\sum_{g}\sigma_1^g(z)  d^2x d^2y d^2z d^2u
$$
Taking for example
$a=c=g ; b=f ; d=e$, we obtain
$$
(g_{12}^0)^2 g_2^0 C_3^3(r,\epsilon_1,\eps_2)\int\displaystyle\sum_{a=1}^M \sigma_1^a(z) d^2z
$$
where
\beqa
 C_3^3(r,\epsilon_1) &=& 8N(N-1) \times \\
 \int\limits_{|y-x|,|z-x|,|w-x| <r}&&\!\!\!\!\!\!\!\!\!\!\!\!\!\!
<\sigma_1(x) \varepsilon_1(y) \varepsilon_1(z) \sigma_1(\infty)>_0
<\varepsilon_2(y)  \varepsilon_2(w)>_0<\varepsilon_2(w)  \varepsilon_2(z)>_0 d^2y
d^2z d^2w \nn
\eeqa
and leads in the Coulomb gas representation to
$$
 16N(N-1) {\cal N} \pi \int
|w|^{-2-3\epsilon_2} |w-1|^{-2-3\epsilon_2} d^2w\int\limits_{y<r}\left (dy\over
y^{1+3\epsilon_1+6\eps_2}\right)
$$
\beq
\label{ic33}
\int |z|^{4\alpha_{12}\alpha_{\overline{k,k-1}}}
|z-1|^{-8\Delta'_{12}+4\alpha^2_{12}+2}
|u|^{4\alpha_+ \alpha_{\overline{k,k-1}}} |u-1|^{4\alpha_+ \alpha_{12}}
|u-z|^{4\alpha_+ \alpha_{12}} d^2z d^2u~.
\eeq
with $4\Delta'_{12}=-2-3\eps_2$.
The third integral is computed in appendix E (see (\ref{res7})) and the final result is

\beq
C_3^3(r,\epsilon_1,\eps_2) =
 -8N(N-1)\pi^3 (g_{12}^0)^2 (g_2^0) {r^{-3\eps_1-6\eps_2}\over 3\eps_1+6\eps_2 }\left[1+{\eps_1\over 2\eps_2}(1+{8\over 3}(2-\lambda_1){\cal R})\right]
\eeq

When we gather all these second and third contributions, we find the renormalisations of $h_i^0$ (\ref{champs1},\ref{champs2}).

\section{Beta function in the disordered case}
\setcounter{equation}{0}
In this appendix, we only give the renormalisation associated to the coupling constants
of one Ising model plus one Potts model with disorder, namely $g_{12}, \Delta_1,\Delta_2,\D_{12}$. We do not give the details of all contributions but only the final results since the integrals we compute are similar to appendix A. The only difference lays in symmetry factors. The results extend to $M$ Ising models coupled to $N$ Potts models with disorder, nevertheless there are much more terms that will simplify when we use the renormalised coupling constants. In the following equations, $n$ represents the replica number.
\beqa
\D_1(r)&=&r^{-3\eps_1}\left[\D_1^0-4\pi(n-2)(\D_1^0)^2{r^{-3\eps_1}\over 3\eps_1}
-4\pi n(\D_{12}^0)^2{r^{-3\eps_2}\over 3\eps_2}-8\pi g_{12}^0\D_{12}^0{r^{-3\eps_2}\over 3\eps_2}\right.\nn\\&&~~~~~~~
+4(n-2)(n-3)(\D_1^0)^3 {\cal I}_{11}+4(n-2) (\D_1^0)^3 {\cal J}_{11}\nn\\&&~~~~~~~
+8(n-2)^2 \D_1^0 (\D_{12}^0)^2  {\cal I}_{12}+4(n-1)\D_1^0 (\D_{12}^0)^2 {\cal J}_{12}\\&&~~~~~~~
+16(n-2) \D_1^0 \D_{12}^0 (\D_{12}^0+g_{12}^0)  {\cal I}_{12}+4 (\D_{12}^0+g_{12}^0)^2\D_1^0{\cal J}_{12}\nn\\&&~~~~~~~
+4\left.\left((2n^2-3n+3) (\D_{12}^0)^2 \D_2^0+2(n-2)(g_{12}^0+\D_{12}^0)\D_{12}^0 \D_2^0
+(g_{12}^0+\D_{12}^0)^2\D_2^0\right){\cal I}_{22}\right]\nn
\eeqa

\beqa
\D_2(r)&=&r^{-3\eps_2}\left[\D_2^0-4\pi(n-2)(\D_2^0)^2{r^{-3\eps_2}\over 3\eps_2}
-4\pi n(\D_{12}^0)^2{r^{-3\eps_1}\over 3\eps_1}-8\pi g_{12}^0\D_{12}^0{r^{-3\eps_1}\over 3\eps_1}\right.\nn\\&&~~~~~~~
+4(n-2)(n-3)(\D_2^0)^3 {\cal I}_{22}+4(n-2) (\D_2^0)^3 {\cal J}_{22}\nn\\&&~~~~~~~
+8(n-2)^2 \D_2^0 (\D_{12}^0)^2  {\cal I}_{21}+4(n-1)\D_2^0 (\D_{12}^0)^2 {\cal J}_{21}\\&&~~~~~~~
+16(n-2) \D_2^0 \D_{12}^0 (\D_{12}^0+g_{12}^0)  {\cal I}_{21}+4 (\D_{12}^0+g_{12}^0)^2\D_2^0{\cal J}_{21}\nn\\&&~~~~~~~
+4\left.\left((2n^2-3n+3) (\D_{12}^0)^2 \D_1^0+2(n-2)(g_{12}^0+\D_{12}^0)\D_{12}^0 \D_1^0
+(g_{12}^0+\D_{12}^0)^2\D_1^0\right){\cal I}_{11}\right]\nn
\eeqa
\beqa
(\D_{12}+g_{12})(r)&=&r^{- {3\over 2}(\eps_1+\eps_2)}\left[(\D_{12}^0+g_{12}^0)-4\pi(n-1)\D_1^0\D_{12}^0{r^{-3\eps_1}\over 3\eps_1}-4\pi(n-1)\D_2^0\D_{12}^0{r^{-3\eps_2}\over 3\eps_2}\right.\nn\\&&
+4(n-1)(n-2)(\D_1^0)^2\D_{12}^0{\cal I}_{11}+4(n-1)(n-2)(\D_2^0)^2\D_{12}^0{\cal I}_{22}\nn\\&&
+2(n-1)(\D_1^0)^2(\D_{12}^0+g_{12}^0){\cal J}_{11}+2(n-1)(\D_2^0)^2(\D_{12}^0+g_{12}^0){\cal J}_{22}\nn\\&&
+4(n-1)(n-2)\D_1^0\D_2^0\D_{12}^0{\cal I}_{12}+4(n-1)\D_1^0\D_2^0(\D_{12}^0+g_{12}^0){\cal I}_{12}\\&&
+4(n-1)(n-2)(\D_{12}^0)^3{\cal I}_{12}+4(n-1)(\D_{12}^0)^2(\D_{12}^0+g_{12}^0)
{\cal I}_{12}\nn\\&& \left.+2(n-1)(\D_{12}^0)^2(\D_{12}^0+g_{12}^0)({\cal J}_{12}+{\cal J}_{21})\right]\nn
\eeqa
\beqa
\D_{12}(r)&=&r^{- {3\over 2}(\eps_1+\eps_2)}\left[
\D_{12}^0-4\pi(n-2)\D_1^0\D_{12}^0{r^{-3\eps_1}\over 3\eps_1}-4\pi(n-2)\D_2^0\D_{12}^0{r^{-3\eps_2}\over 3\eps_2}\right.\nn\\&&
-4\pi\D_1^0(\D_{12}^0+g_{12}^0){r^{-3\eps_1}\over 3\eps_1}
-4\pi\D_2^0(\D_{12}^0+g_{12}^0){r^{-3\eps_2}\over 3\eps_2}\nn\\&&
+4(n-2)^2(\D_1^0)^2\D_{12}^0{\cal I}_{11}+4(n-2)(\D_1^0)^2(\D_{12}^0+g_{12}^0){\cal I}_{11}
+2(n-1)(\D_1^0)^2\D_{12}^0{\cal J}_{11}\nn\\&&
+4(n-2)^2(\D_2^0)^2\D_{12}^0{\cal I}_{22}+4(n-2)(\D_2^0)^2(\D_{12}^0+g_{12}^0){\cal I}_{22}
+2(n-1)(\D_2^0)^2\D_{12}^0{\cal J}_{22}\nn\\&&
+4(n-1)^2\D_1^0\D_2^0\D_{12}^0{\cal I}_{12}+4(n-2)\D_1^0\D_2^0g_{12}^0{\cal I}_{12}\\&&
+4(n-2)(n-3)(\D_{12}^0)^3{\cal I}_{12}+2(n-2)(\D_{12}^0)^3({\cal J}_{12}+{\cal J}_{21})\nn\\&&
+12(n-2)(\D_{12}^0)^2(\D_{12}^0+g_{12}^0){\cal I}_{12}+4(\D_{12}^0+g_{12}^0)^2\D_{12}^0{\cal I}_{12}\nn\\&&\left.+2(\D_{12}^0+g_{12}^0)^2\D_{12}^0(
{\cal J}_{12}+{\cal J}_{21})\right]\nn
\eeqa
In this system of equations, we have defined for readability the following integrals
\beq
{\cal I}_{11}=4\pi^2{r^{-6\eps_1}\over 9\eps_1^2}~~;~~
{\cal I}_{22}=4\pi^2{r^{-6\eps_2}\over 9\eps_2^2}
\eeq
\beq
{\cal I}_{12}={\cal I}_{21}=4\pi^2{r^{-3(\eps_1+\eps_2)}\over 9\eps_1\eps_2}
\eeq
\beq
{\cal J}_{11}=2\pi^2{r^{-6\eps_1}\over 3\eps_1}\left[1+{2\over 3\eps_1}\right]~~;~~
{\cal J}_{22}=2\pi^2{r^{-6\eps_2}\over 3\eps_2}\left[1+{2\over 3\eps_2}\right]
\eeq
\beq
{\cal J}_{12}=4\pi^2{r^{-3(\eps_1+\eps_2)}\over 3(\eps_1+\eps_2)}\left[1+{2\over 3\eps_1}\right]~~;~~
{\cal J}_{21}=4\pi^2{r^{-3(\eps_1+\eps_2)}\over 3(\eps_1+\eps_2)}\left[1+{2\over 3\eps_2}\right]
\eeq
The calculations of the last four integrals is recalled in the appendix E. The last step consists in deriving the beta function, $r$ being the scale variable.

\section{Computation of integrals}
\setcounter{equation}{0}
In this appendix, we detail the calculation of some integrals we have used in the previous appendices. These integrals take the general form
\beq
I =
\int |x|^{2a}  |x-1|^{2b}
|y|^{2a'} |y-1|^{2b'}
|x-y|^{4g} d^2x d^2y
\eeq
where the integration is performed over the whole complex plane.
Some general techniques to compute this kind of integrals have been shown in  ref.\cite{fateev, dots3}.  We just recall the main steps in this appendix (see also ref. \cite{dots2} for the method). This
integral can be  decomposed into holomorphic and antiholomorphic parts~:
\beq
\label{int}
I = s(b) s(b') \left[ J_1^+  J_1^- + J_2^+  J_2^-\right]
+ s(b) s(b'+2g) J_1^+ J_2^- + s(b+2g) s(b') J_2^+ J_1^-
\eeq
where $s(x)$ corresponds to $sin(\pi x)$ and
\beqa
\label{jis}
J_1^+ &=& J(a,b,a',b',g)~;~J_2^+ = J(b,a,b',a',g) \nn \\
J_1^- &=& J(b,-2-a-b-2g,b',-2-a'-b'-2g,g) \\
J_2^- &=& J(-2-a-b-2g,b,-2-a'-b'-2g,b',g) \nn
\eeqa
Here, we used the notation
\beqa
\label{hyp}
 J(a,b,a',b',g) =
\int\limits_{0}^{1} du \int\limits_{0}^{1} dv~
u^{a+a'+2g+1} (1-u)^b v^{a'}&&\!\!\!\!\!\!\!\! (1-v)^{2g} (1-uv)^{b'}\\
= {\Gamma(2+a+a'+2g)\Gamma(1+b)\Gamma(1+a')\Gamma(1+2g)\over
\Gamma(3+a+a'+b+2g)\Gamma(2+a'+2g)}
&&\!\!\!\!\!\!\!\!\sum_{k=0}^{\infty}{(-b')_k (2+a+a'+2g)_k (1+a')_k \over
k! (3+a+a'+b+2g)_k (2+a'+2g)_k} \nonumber
\eeqa
and
$$
(a)_k = a(a+1)...(a+k-1)
$$
The $J$ integrals appearing in (\ref{int}) are not all independent. Using
contour deformation of integrals it can be shown that we have the following
relations:
\beq
\label{rel1}
s(2g+a+b)J_1^- + s(a+b)J_2^- = {s(a)\over s(2g+a'+b')}
\left( s(a') J_1^+ + s(2g+a')J_2^+ \right)
\eeq
\beq
\label{rel}
s(2g+a'+b')J_2^- + s(a'+b')J_1^- = {s(a')\over s(2g+a+b)}
\left( s(a) J_2^+ + s(2g+a)J_1^+ \right)
\eeq
With these formulas, the integrals appearing in the appendices A,C can be calculated as a power series of $\epsilon_1$ and $\eps_2$.

\vskip 0.5cm
$\bullet$
Let us detail the integral referred as (\ref{ia22}). This integral has been computed in \cite{dots2} but we recall it since we find some errors in the intermediary steps of ref. \cite{dots2}.

The coefficients
corresponding to that case are
\beq
a=-2a' =2+3\epsilon_1 \quad ; \quad b'=-{1\over 3} -\epsilon_1 \quad ; \quad
b=2g=-{4\over 3} -\epsilon_1 
\eeq
Substituting these values in (\ref{int}), we obtain at the second order in
$\epsilon_1$
\beqa
\label{e1}
I&=&{3\over4}(J_2^+ -J_1^+)(J_1^- - J_2^-) -{\sqrt{3}\pi\epsilon_1\over
4}\left( J_1^+(2J_1^- +J_2^-)+J_2^+(J_1^- +2J_2^-)\right)\nn\\&&
+{\pi^2\eps_1^2\over 8}\left( 4(J_1^+J_1^-+J_2^+J_2^-)-19(J_1^+J_2^-+J_2^+J_1^-)\right)
\eeqa
Now we write the $J_i^-$ as functions of the $J_i^+$ with the help of
relations (\ref{rel1},\ref{rel})
\beqa
J_1^-&=&-J_1^++2\left(1-{\pi\sqrt{3}\eps_1\over 2}-{\pi^2\eps_1^2\over 4}\right)J_2^+\\
J_2^-&=&-\left(1+\pi\sqrt{3}\eps_1+{19\over 2}\pi^2\eps_1^2\right)J_1^++2\left(1+{3\pi\sqrt3\eps_1\over 2}+{33\pi^2\eps_1^2\over 4}\right)J_2^+
\eeqa
 The equation (\ref{e1})  simplifies to
\beq
I=3\sqrt{3}\pi\epsilon_1 J_1^+ J_2^+ -{9\over 2}\pi^2\epsilon_1^2 (J_1^+)^2
\eeq
The $J_1^+$ and $J_2^+$ can now be computed explicitly using formulas
(\ref{jis},\ref{hyp}). We obtain~:
\beqa
J_1^+ &=&{4\over 9\epsilon_1}{\Gamma^2(-{1\over3})\over
\Gamma({1\over 3})} + O(cst) \nn\\
J_2^+ &=&{2\over 9} ({3\over 2}+\pi\sqrt{3}){\Gamma^2(-{1\over3})\over
\Gamma({1\over 3})} + O(\epsilon) \nn
\eeqa
The final result is therefore (at lowest order in $\eps_1$)
\beq
\label{res1}
I= \sqrt{3}\pi{\Gamma^4(-{1\over3})\over
\Gamma^2(-{2\over 3})}
\eeq

$\bullet$
We now compute the integral (\ref{ia42}). The coefficients are now
\beq
a=-2a' =2+3\epsilon_1 \quad ; \quad b'=-{1\over 3} +{1\over 2}(\epsilon_1-3\eps_2) \quad ; \quad
b=2g=-{4\over 3} -\epsilon_1 
\eeq
Substituting these values in (\ref{int}), we obtain at the second order in
$\epsilon$
\beqa
\label{e2}
I&=&{3\over4}(J_2^+ -J_1^+)(J_1^- - J_2^-)\nn\\&& -{\sqrt{3}\pi\epsilon_1\over
8}\left( J_1^+J_1^- +J_2^+J_2^- -J_1^+J_2^- +5J_2^+J_1^-\right)\nn\\&&
 -{3\sqrt{3}\pi\epsilon_2\over
8}\left( J_1^+J_1^- +J_2^+J_2^- +J_1^+J_2^- -J_2^+J_1^-\right)\nn\\&&
+{\pi^2\eps_1^2\over 32}\left( 19(J_1^+J_1^-+J_2^+J_2^-)-19J_1^+J_2^--43J_2^+J_1^-)\right)
\\&&
-{3\pi^2\eps_1\eps_2\over 16}\left( 5(J_1^+J_1^-+J_2^+J_2^-)+5J_1^+J_2^-+J_2^+J_1^-)\right)\nn\\&&
+{27\pi^2\eps_2^2\over 32}\left( J_1^+J_1^-+J_2^+J_2^- -J_1^+J_2^- -J_2^+J_1^-)\right)\nn 
\eeqa
Now we write the $J_i^-$ as functions of the $J_i^+$ with the help of
relations (\ref{rel1},\ref{rel}) and
the equation (\ref{e2})  simplifies after tedious algebra to
\beq
I=3\sqrt{3}\pi{\epsilon_1^2\over \eps_2} J_1^+ J_2^+ -{3\sqrt{3}\over 4}\pi\eps_1^2[{1\over \eps_2}-{1\over \eps_1}](J_1^+)^2 -{3\over 2}\pi^2\epsilon_1^2 (J_1^+)^2[1+2{\eps_1\over \eps_2}]
\eeq
The $J_1^+$ and $J_2^+$ are now be computed explicitly using formulas
(\ref{jis},\ref{hyp}). We find~:
\beqa
J_1^+ &=&{4\over 9\epsilon_1}{\Gamma^2(-{1\over3})\over
\Gamma({1\over 3})}\left(1 + ({3\over 2}+{\pi\over \sqrt{3}})\eps_1+o(\eps_i)\right) \nn\\
J_2^+ &=&{2\over 9} ({3\over 2}+\pi\sqrt{3}){\Gamma^2(-{1\over3})\over
\Gamma({1\over 3})} + O(\epsilon_i) \nn
\eeqa
The calculation of the sums (\ref{hyp}) is quite heavy. We refer to ref. \cite{dots3} where some tricks concerning their manipulations are given.

The final result is therefore 
\beq
\label{res2}
I= \sqrt{3}\pi{\Gamma^4(-{1\over3})\over
\Gamma^2(-{2\over 3})}\left(1+{1\over 3\eps_1}-{1\over 3\eps_2}\right)
\eeq
Note that we recover the result (\ref{res1}) when $\eps_1=\eps_2$ as it should be. During the calculation, we have checked that the terms in ${\eps_1\over \eps_2}$ simplifiy each other in order to ensure renormalisability. Nevertheless, we are obliged to admit the result for terms in  ${\eps_1^p\over \eps_2}$. On the other hand, singurities in ${1\over \eps_i}$ are necessary in order to compensate the ones coming from one loop calculations.

\vskip 0.5cm

$\bullet$
 We now compute the integral (\ref{ic12}). Here, we have~:
\beq
a=-{1\over3} + \lambda_1\epsilon_1 =-2 a' \quad ; \quad
b = 2g = b'-1 = -{4\over3} - \epsilon_1
\eeq
Substituting these values, we find
\beq
\label{e3}
I={3\over4}(J_2^+ -J_1^+)(J_1^- - J_2^-) -{\sqrt{3}\pi\epsilon_1\over
4}\left( J_1^+(2J_1^- +J_2^-)+J_2^+(J_1^- +2J_2^-)\right)
\eeq
Then, relating the $J_i^-$ to the $J_i^+$,
\beqa
J_1^+&=&-\pi\sqrt{3}\eps_1(2-\lambda_1)J_1^--(1-\pi\sqrt{3}\eps_1(2-\lambda_1))J_2^-\\
J_2^+&=&-\pi{\sqrt{3}\over 3}\eps_1(2-\lambda_1)J_1^--(1+\pi{\sqrt{3}\over 3}\eps_1(1+\lambda_1))J_2^-
\eeqa
 we obtain for $I$~:
\beq
I = {3\sqrt{3}\pi \epsilon_1\over 2}\left(J_2^-\right)^2 +
 {\sqrt{3} (2-\lambda_1)\pi \epsilon_1\over 2}\left(J_2^- -J_1^- \right)^2
+ O(\epsilon_1^2)
\eeq
These $J_i^-$ can again be computed with the help of
(\ref{jis},\ref{hyp})~:
\beq
\label{jint}
J_1^- = J_2^- + {\Gamma(-{1\over3})\Gamma({1\over6})\over
\Gamma(-{1\over6})} = {1\over2} {\Gamma^2(-{1\over3})\over
\Gamma(-{2\over3})} +  {\Gamma(-{1\over3})\Gamma({1\over6})\over
\Gamma(-{1\over6})}+ O(\epsilon_1)
\eeq
Therefore~:
\beq
\label{res3}
I = {3\sqrt{3} \pi \epsilon_1\over 8} {\Gamma^4(-{1\over3})\over
\Gamma^2(-{2\over3})}\left[1 + {4\over 3} (2-\lambda_1)\pi \epsilon_1
{\Gamma^2(-{2\over3})\Gamma^2({1\over6})\over
\Gamma^2(-{1\over3})\Gamma^2(-{1\over6})}\right]+ O(\epsilon_1^2)
\eeq

\vskip 0.5cm
$\bullet$
 We now compute the integral (\ref{ic22}). Here, we have now~:
\beq
a=-{1\over3} + \lambda_1\epsilon_1 =-2 a' \quad ; \quad
b = 2g = -{4\over3} - \epsilon_1\quad ; \quad b'=-{1\over 3}+{1\over 2}(\eps_1-3\eps_2)
\eeq
Substituting these values, we find
\beqa
I&=&{3\over4}(J_2^+ -J_1^+)(J_1^- - J_2^-)\nn\\&& -{\sqrt{3}\pi\epsilon_1\over
8}\left( J_1^+J_1^- +J_2^+J_2^- -J_1^+J_2^- +5J_2^+J_1^-\right)\\&&
 -{3\sqrt{3}\pi\epsilon_2\over
8}\left( J_1^+J_1^- +J_2^+J_2^- +J_1^+J_2^- -J_2^+J_1^-\right)\nn
\eeqa
The relations between the $J_i^-$ to the $J_i^+$ reamin the same as above
and
 we obtain for $I$~:
\beq
I = {3\sqrt{3}\pi (\epsilon_1+\eps_2)\over 4}\left(J_2^-\right)^2 +
 {\sqrt{3} (2-\lambda_1)\pi \epsilon_1\over 2}\left(J_2^- -J_1^- \right)^2
+ O(\epsilon_1^2)
\eeq
The values of the $J_i^-$ remain the same  as (\ref{jint}) and thus
\beq
\label{res4}
I = {3\sqrt{3} \pi \epsilon_1\over 16} {\Gamma^4(-{1\over3})\over
\Gamma^2(-{2\over3})}
\left[\eps_2+\eps_1\left(1+{8\over 3}(2-\lambda_1){\Gamma^2(-{2\over3})\Gamma^2({1\over6})\over
\Gamma^2(-{1\over3})\Gamma^2(-{1\over6})}\right)\right]
\eeq

\vskip 0.5cm
$\bullet$
We now calculate the integral (\ref{ic13}).
We have the following coefficients:
\beq
a=-{1\over3} + \lambda\epsilon_1 =-2 a' \quad ; \quad
b = 2g =  -{4\over3} - \epsilon_1 \quad ; \quad
b'=-{1\over3}-{5\over2}\epsilon_1
\eeq
With similar manipulations, the integral reads
\beq
I = {9\sqrt{3}\pi \epsilon_1\over 4}\left(J_2^-\right)^2 +
 {\sqrt{3} (2-\lambda_1)\pi \epsilon_1\over 2}\left(J_2^- -J_1^- \right)^2
+ O(\epsilon_1^2)~,
\eeq
and the final result is
\beq
\label{res5}
I = {9\sqrt{3} \pi \epsilon_1\over 16} {\Gamma^4(-{1\over3})\over
\Gamma^2(-{2\over3})}\left[1 + {8 \over 9}(2-\lambda_1)\pi \epsilon_1
{\Gamma^2(-{2\over3})\Gamma^2({1\over6})\over
\Gamma^2(-{1\over3})\Gamma^2(-{1\over6})}\right]+ O(\epsilon_1^2)
\eeq

\vskip 0.5cm
$\bullet$
We now compute the integral (\ref{ic23}).
We have:
\beq
a=-{1\over3} + \lambda\epsilon_1 =-2 a' \quad ; \quad
b = 2g =  -{4\over3} - \epsilon_1 \quad ; \quad
b'=-{1\over3}-(\eps_1+{3\over2}\epsilon_2)
\eeq
With similar manipulations, the integral reads
\beq
I = {3\sqrt{3}\pi (2\epsilon_1+\eps_2)\over 4}\left(J_2^-\right)^2 +
 {\sqrt{3} (2-\lambda_1)\pi \epsilon_1\over 2}\left(J_2^- -J_1^- \right)^2
+ O(\epsilon_1^2)~,
\eeq
and the final result is
\beq
\label{res6}
I = {3\sqrt{3} \pi \epsilon_1\over 16} {\Gamma^4(-{1\over3})\over
\Gamma^2(-{2\over3})}\left[\eps_2+2\eps_1\left(1 + {4\over 3}(2-\lambda_1)\pi \epsilon_1
{\Gamma^2(-{2\over3})\Gamma^2({1\over6})\over
\Gamma^2(-{1\over3})\Gamma^2(-{1\over6})}\right)\right]+ O(\epsilon^2)
\eeq

\vskip 0.5cm
$\bullet$
Concerning the integral (\ref{ic33}),
we have the following coefficients:
\beq
a=-{1\over3} + \lambda\epsilon_1 =-2 a' \quad ; \quad
b = 2g =  -{4\over3} - \epsilon_1 \quad ; \quad
b'=-{1\over3}+{\eps_1\over 2}-3\epsilon_2
\eeq
With similar manipulations, the integral reads
\beq
I = {3\sqrt{3}\pi (\epsilon_1+2\eps_2)\over 4}\left(J_2^-\right)^2 +
 {\sqrt{3} (2-\lambda_1)\pi \epsilon_1\over 2}\left(J_2^- -J_1^- \right)^2
+ O(\epsilon_1^2)~,
\eeq
and the final result is
\beq
\label{res7}
I = {3\sqrt{3} \pi \epsilon_1\over 16} {\Gamma^4(-{1\over3})\over
\Gamma^2(-{2\over3})}\left[2\eps_2+\eps_1\left(1 + {8\over 3}(2-\lambda_1)\pi \epsilon_1
{\Gamma^2(-{2\over3})\Gamma^2({1\over6})\over
\Gamma^2(-{1\over3})\Gamma^2(-{1\over6})}\right)\right]+ O(\epsilon^2)
\eeq

\eject
 
\eject

\vskip 0.5 truecm
\begin{center}
{\bf CAPTIONS}
\end{center}

\vskip 1.5cm
\no
FIG.1 : The projection of the flow in the $(g_{12},g_2)$ plane for one Ising model coupled to three Potts models. Two new tricritical points denoted by $T_1,~T_2$ are found.

\vskip 1 truecm
\no
FIG.2 : The projection of the flow in the $(g_{1},g_{12})$ plane for two Ising models coupled to  a Potts model. The point $P$ separates the fixed point line $g_1$ in  stable  and unstable parts.

\vskip 1 truecm
\no
FIG.3 : The projection of the flow in the $(g_{12},\D_{12})$ plane for one Ising model coupled to  one Potts model under weak disorder. We clearly see that disorder prevents the flow from running in a strong coupling regime and therefore makes the models decouple.

\vskip 1 truecm
\eject
\begin{figure}
\epsfxsize=14cm
$$
\epsfbox{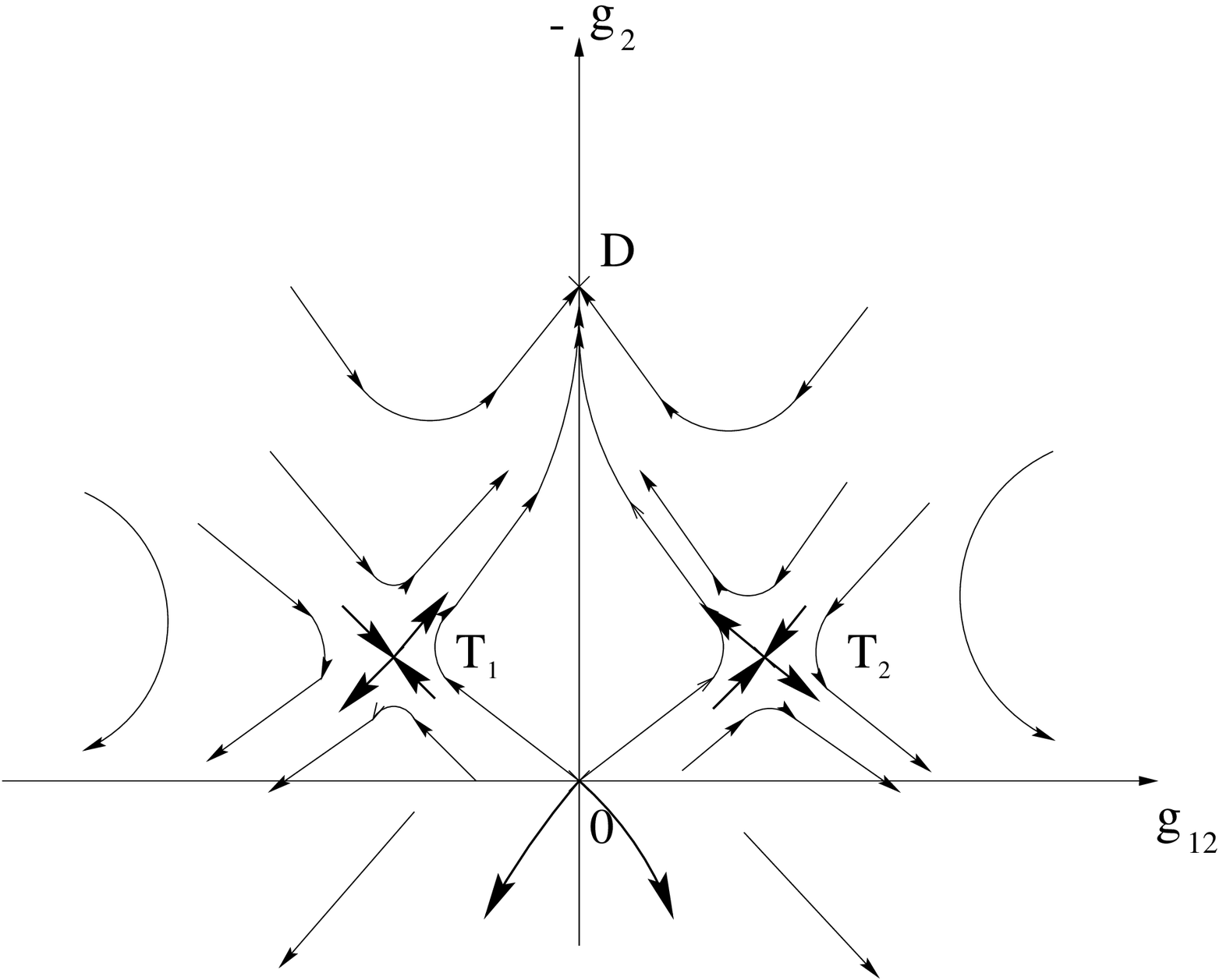}
$$
{\bf Figure 1}
\end{figure}

\eject

\begin{figure}
\epsfxsize=14cm
$$
\epsfbox{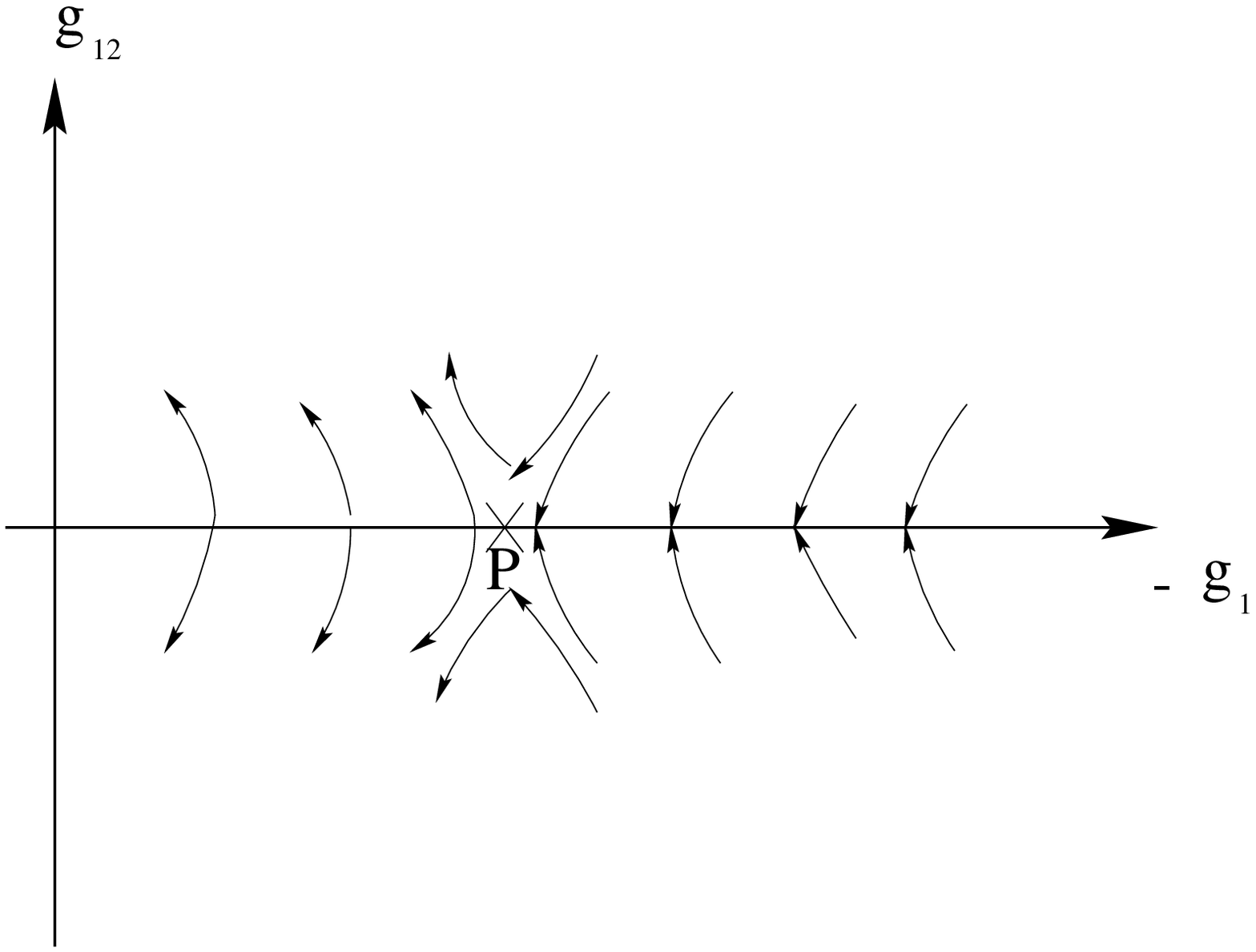}
$$
{\bf Figure 2}
\end{figure}

\newpage
\begin{figure}
\psfig{figure=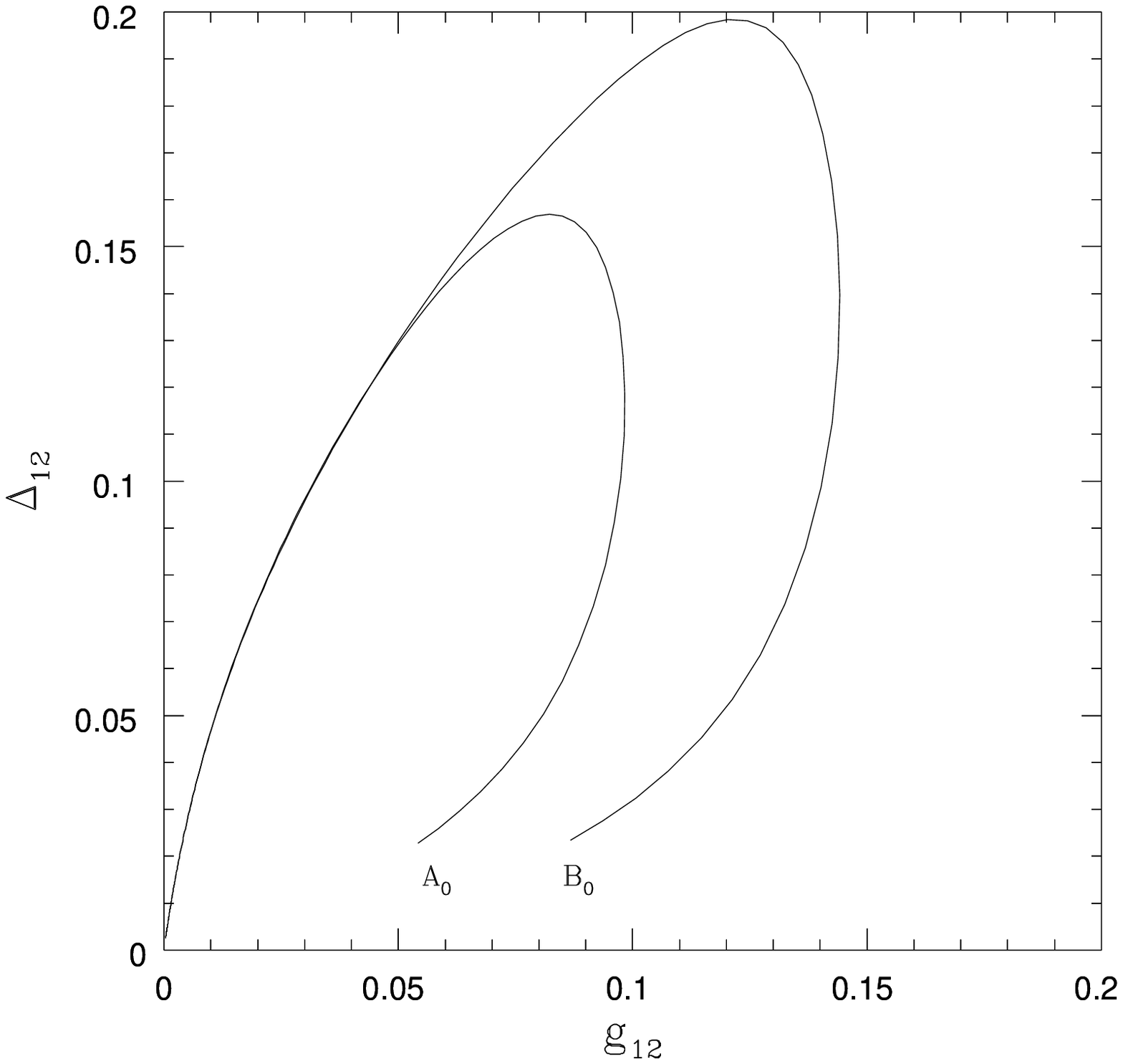,height=14cm,width=18cm}
{\bf Figure 3}
\end{figure}

\end{document}